\documentclass[aps,prb,english,amsmath,amsfonts,amssymb,superscriptaddress,twocolumn,showpacs,citeautoscript,reprint,longbibliography]{revtex4-1}
\usepackage[T1]{fontenc}
\usepackage[latin9]{inputenc}
\usepackage{amsmath}
\usepackage{amssymb}
\usepackage{wasysym}
\usepackage{adjustbox}
\usepackage{esint}
\usepackage{graphicx}
\usepackage{times}
\usepackage{nicefrac}
\usepackage{appendix}
\usepackage{multirow}
\usepackage{textcase}
\usepackage{subfigure}
\usepackage{empheq}
\usepackage{color}
\usepackage{leftidx}
\usepackage{hhline}
\usepackage{makecell}
\usepackage{stackrel}
\usepackage{lineno}
\makeatletter
\@ifundefined{textcolor}{}
{%
 \definecolor{BLACK}{gray}{0}
 \definecolor{WHITE}{gray}{1}
 \definecolor{RED}{rgb}{1,0,0}
 \definecolor{GREEN}{rgb}{0,1,0}
 \definecolor{BLUE}{rgb}{0,0,1}
 \definecolor{CYAN}{cmyk}{1,0,0,0}
 \definecolor{MAGENTA}{cmyk}{0,1,0,0}
 \definecolor{YELLOW}{cmyk}{0,0,1,0}
}

\renewcommand{\vec}[1]{\mathbf{#1}}

\renewcommand{\Im}{\operatorname{Im}}

\newcommand{\asym}{\operatorname{asym}}

\renewcommand{\b}{\beta}

\newcommand{\add}[1]{\if\a\b{{\color{red} #1}}\else{#1}\fi}

\newcommand{\bracket}[1]{\langle #1 \rangle}
\newcommand{\ket}[1]{| #1 \rangle}
\newcommand{\bra}[1]{\langle #1 |}
\newcommand{\im}{\operatorname{i}}

\renewcommand{\eqref}[1]{(\ref{eq:#1})}

\newcommand{\Eqref}[1]{Equation~\ref{eq:#1}}
\newcommand{\figref}[1]{Fig.~\ref{fig:#1}}

\newcommand{\Secref}[1]{Section~\ref{sec:#1}}

\newcommand{\trace}[1]{{\rm Tr} \left[ #1 \right]}

\newcommand{\VV}{\mathbb{V}}
\newcommand{\GG}{\mathbb{G}}

\newcommand{\PP}{\mathbb{P}}
\newcommand{\Gvac}{\mathbb{G}^{\mathrm{vac}}}

\makeatother
\usepackage{babel}

\begin{document}
\title{Mechanical relations between conductive and radiative heat transfer}

\author{Prashanth S. Venkataram}
\affiliation{Department of Electrical Engineering, Princeton
  University, Princeton, New Jersey 08544, USA}

\author{Riccardo Messina}
\affiliation{Laboratoire Charles Fabry, UMR 8501, Institut d'Optique,
  CNRS, Universit\'e Paris-Sud 11, 2, Avenue Augustin Fresnel, 91127
  Palaiseau Cedex, France}

\author{Juan Carlos Cuevas}
\affiliation{Departamento de F\'{\i}sica Te\'orica de la Materia
  Condensada and Condensed Matter Physics Center (IFIMAC), Universidad
  Aut\'onoma de Madrid, E-28049 Madrid, Spain}

\author{Philippe Ben-Abdallah}
\affiliation{Laboratoire Charles Fabry, UMR 8501, Institut d'Optique,
  CNRS, Universit\'e Paris-Sud 11, 2, Avenue Augustin Fresnel, 91127
  Palaiseau Cedex, France}

\author{Alejandro W. Rodriguez}
\affiliation{Department of Electrical Engineering, Princeton
  University, Princeton, New Jersey 08544, USA}

\date{\today}

%% Recent experimental advances have probed phonon conduction and
%% radiative heat transfer at separations close to contact, but have
%% led to conflicting claims about the validity of prevailing
%% theoretical models for phonon conductive heat transfer and
%% radiative heat transfer in such regimes.
\begin{abstract}
We present a general nonequilibrium Green's function formalism for
modeling heat transfer in systems characterized by linear response
that establishes the formal algebraic relationships between phonon and
radiative conduction, and reveals how upper bounds for the former can
also be applied to the latter. We also propose an extension of this
formalism to treat systems susceptible to the interplay of conductive
and radiative heat transfer, which becomes relevant in atomic systems
and at nanometric and smaller separations where theoretical
descriptions which treat each phenomenon separately may be
insufficient. We illustrate the need for such coupled descriptions by
providing predictions for a low-dimensional system of carbyne wires in
which the total heat transfer can differ from the sum of its radiative
and conductive contributions. Our framework has ramifications for
understanding heat transfer between large bodies that may approach
direct contact with each other or that may be coupled by atomic,
molecular, or interfacial film junctions.
\end{abstract}

\maketitle 

Characterizing radiative and conductive heat transfer at the nanoscale
is essential to understanding the operation of a wide variety of
systems and technologies, including heat sinks, thermoelectric
devices, thermal microscopy, thermal magnetic recording devices,
coherent thermal sources, optoelectronic and optomechanical devices,
and thermophotovoltaic devices~\cite{VolokitinPRB2001, SegalARPC2016,
  SongAIPADV2015, CuevasACSPHOTONICS2018, TianPRB2012, TianPRB2014,
  BurklePRB2015, KlocknerPRB2017, PolimeridisPRB2015,
  LenertNATURENANO2014}. Much progress has been made toward
experimentally measuring heat conduction by phonons in molecular
junctions and interfaces at contact~\cite{LeeAPL1997, LuoPCCP2013,
  LuckyanovaSCIENCE2012, CuiSCIENCE2017, CuiNATURE2019,
  MossoNATURE2017, MossoNANOLETT2019}, as well as radiative heat
transfer between objects at separations $\gtrsim
10~\mathrm{nm}$~\cite{ShenNANOLETT2009, GuhaNANOLETT2012,
  StGelaisNANOLETT2014, CahillAPR2014, CuevasACSPHOTONICS2018,
  SongNATURENANO2015, StGelaisNATURE2016, OttensPRL2011}. Most
commonly, conduction in the linear response regime is described
atomistically using the nonequilibrium Green's function (NEGF)
method~\cite{SegalARPC2016, MingoPRB2003, TianPRB2012, TianPRB2014,
  DharJSP2006, KlocknerPRB2017, BurklePRB2015, KlocknerPRB2016,
  KlocknerPRB2018, SadasivamPRB2017, DubiRMP2009}, while radiative
heat transfer is modeled through continuum fluctuational
electrodynamics~\cite{OteyJQSRT2014, PolimeridisPRB2015,
  RodriguezPRB2013, KrugerPRB2012, VolokitinRMP2007,
  CuevasACSPHOTONICS2018}. However, recent
experiments~\cite{KimNATURE2015, KloppstechNATURE2017, CuiNATURE2017,
  FongNATURE2019} have yielded conflicting accounts of the nature of
heat transfer in the extreme near-field (ranging from subnanometric
separations to $\lesssim 10~\mathrm{nm}$), raising questions about the
interplay between conduction and radiation at such small
separations. Simultaneously, recent theoretical
works~\cite{YuNATURE2017, JiangPRB2017, ChiloyanNATURE2015,
  ZhangPRB2018, WangPRE2018, DominguesPRL2005, TangPRAPP2019,
  KlocknerPRB2017} have begun to shed light on the connections between
the formalisms of conductive (whether electronic or phononic) and
radiative heat transfer, but these have typically been subject to
restrictions including neglect of electromagnetic retardation and
consideration of translationally symmetric systems like planar sheets
or slabs. In this paper, we present a unified linear response
formalism that can describe phonon conductive heat transfer (PCHT) and
radiative heat transfer (RHT) for arbitrary geometries and
separations. The approach puts descriptions of both effects on the
same algebraic footing, which is useful for drawing mathematical and
physical analogies. For illustration, we demonstrate that recent
analytical upper limits on PCHT can be applied to RHT, and further
show that our framework can be used to describe situations where both
effects couple and contribute significantly to net heat transfer, of
particular relevance to recent and ongoing experiments at the
nanoscale~\cite{CahillAPR2014, CuevasACSPHOTONICS2018, KimNATURE2015,
  KloppstechNATURE2017, CuiNATURE2017, FongNATURE2019}.

Nanoscale PCHT has thus far been treated through atomistic theoretical
frameworks primarily using one of two classes of methods. One approach
is the so-called NEGF method~\cite{SegalARPC2016, MingoPRB2003,
  TianPRB2012, TianPRB2014, DharJSP2006, KlocknerPRB2017,
  BurklePRB2015, KlocknerPRB2016, KlocknerPRB2018, SadasivamPRB2017,
  DubiRMP2009}, typically used to model heat transfer between two
large or semi-infinite metallic or polar dielectric leads across a
junction, taken to be either a single atom or molecule or a thin
interfacial film; this method has not been applied so much to smaller
material bodies exchanging heat. The NEGF method models each material
body as being made of atoms, each of which corresponds to harmonic
oscillator degrees of freedom along each Cartesian direction
representing chemical bonds between neighboring atoms, whose strengths
are typically computed via density functional theory. This harmonic
model is a frequency domain method, and is valid at temperatures
$\lesssim 500~\mathrm{K}$, when the spatial dimensions relevant to
energy transport between the bodies under consideration are smaller
than the phonon mean free path in the material, and when other tunable
anharmonicities are negligible~\cite{SegalARPC2016, SadasivamPRB2017,
  KlocknerPRB2017}. It is this NEGF method that we use to treat PCHT
in this work, which is why we consistently use the term PCHT to
specifically refer to \emph{coherent} thermal phonon transport in the
\emph{linear} regime under the aforementioned conditions. Another
typical approach for modeling PCHT is based on molecular
dynamics~\cite{CuiJPCA2015, HenryPRL2008, EsfarjaniPRB2011,
  NoyaPRB2004, DubiRMP2009}, which is a time domain method that
captures anharmonicity in short- and long-range interactions but
frequently requires complicated empirical functional forms for
interaction potentials.

RHT is typically treated using fluctuational electrodynamics, in which
material bodies are modeled to have continuum susceptibilities that
respond to EM fields propagating between them. Recent analytical and
computational formulations include discrete dipolar and multipolar
methods~\cite{EdalatpourPRB2016, EdalatpourJQSRT2016,
  EdalatpourJQSRT2014, PerezMadridPRB2008, MessinaPRB2013}, scattering
matrix methods~\cite{KrugerPRB2012, VolokitinPRB2001,
  VolokitinRMP2007, BimontePRA2009, MessinaPRA2011, MessinaPRB2017},
finite-difference time domain techniques~\cite{LuoPRL2004,
  RodriguezPRL2011, OteyJQSRT2014}, and surface or volume integral
equation methods~\cite{RodriguezPRB2013, PolimeridisPRB2015,
  ReidPROCIEEE2013}. With the exception of finite-difference time
domain methods, all of the other methods discussed are frequency
domain methods, which require linear media; however, unlike the case
of PCHT, the assumption of linear response is valid under a much
broader range of scenarios (including temperature ranges) of relevance
to RHT. These methods capture long-range EM effects, but their
tendency to use semi-empirical rather than ab-initio calculations
makes them best suited for separations $\gtrsim 10~\mathrm{nm}$, and
their typical neglect of nonlocality and boundary effects at the
atomic scale can lead to unphysical predictions as the objects
undergoing RHT approach contact. The fact that current experiments are
beginning to probe smaller systems~\cite{KimNATURE2015,
  KloppstechNATURE2017, CuiNATURE2017} suggests a need for better
understanding the connections between PCHT and RHT at nanoscale and
smaller separations.

Our paper is organized as follows. In~\Secref{genform}, we explain the
general NEGF formalism for computing heat transfer in a system of
massless bosonic excitations exhibiting linear response in different
collections of bosonic degrees of freedom, which we generically call
``components''. We derive Landauer-like formulas for the spectrum of
energy exchange between any two components, either coupled directly to
one another or by a third, and then derive fully general Landauer
bounds on heat transfer from them, decomposing the spectrum into
transmission channels and bounding the transmission in each channel
above by unity. We then explain the relationship between the general
NEGF formalism and its application to PCHT and RHT, described in
\Secref{PCHT} and \Secref{RHT}, respectively. In \Secref{comparison},
we identify the relevant components and their couplings, and further
clarify the analogies between PCHT and RHT, making it abundantly clear
that RHT and PCHT are simply different manifestations of the same
abstract principles of energy transport in linear systems. Beyond
simply highlighting the abstract connections between the formalisms,
in~\Secref{unification}, we apply the general NEGF formalism to
consider PCHT and RHT in a unified manner. We show that far from
overcomplicating matters, such a unification is \emph{necessary} in
certain regimes. In particular, we consider a model system consisting
of collinear atomically thin wires, and show that the resulting net
heat transfer power does not simply follow from the sum of the
individual radiative and conductive contributions, and may in fact
fall below either or both of these contributions. Such an illustration
is made possible by an extension of the retarded many-body framework
of mesoscale fluctuational EM~\cite{VenkataramPRL2018,
  VenkataramPRL2017, VenkataramSCIADV2019}, which can account for
atom-scale features of material response. Concluding remarks are given
in~\Secref{conclusion}.

\section{General linear response NEGF formalism for heat transfer} \label{sec:genform}

Consider a generic system exhibiting generalized displacements labeled
$x(t)$, which may represent electronic wavefunctions, collective
nuclear oscillations giving rise to phonons, EM fields, or other
oscillatory phenomena, and respond linearly to generalized forces
labeled $F(t)$. These degrees of freedom (DOFs) constitute collections
which we generically call ``components''. Each component exhibits
linear equations of motion representing its internal dynamics in
isolation and in response to external forces, and each component may
be linearly coupled to other components leading to energy transport
among them. The following sections will make clear the identities of
the components, couplings, generalized displacements, and generalized
forces in different systems of interest, like PCHT or RHT; this
section focuses on deriving relevant fully general formulas for energy
transport among generic coupled components.

Generically, in the time domain, the power radiated or absorbed by a
component may be written as $\dot{W} = \left\langle F(t)
\frac{\partial x}{\partial t} \right\rangle$. Here, $\bracket{\ldots}$
denotes a time average in the steady-state, which is equivalent to an
ensemble average due to ergodicity. As we have specified that the
internal dynamics and couplings are linear, we may equivalently work
in the frequency domain, making it easier to apply the
fluctuation--dissipation theorem~\cite{Novotny2006} and thereby
replace such ensemble averages with deterministic quantities
representing the dissipation of the system.

In the frequency domain (where we will generally suppress dependence
on angular frequency $\omega$ in the notation), we label these
generalized displacements as $\ket{x}$ and the generalized forces as
$\ket{F}$, as these quantities are vectors in a complex Hilbert space
with the standard inner product. One of the operators relevant to this
Hilbert space are the dynamical operator $\hat{Z}^{(0)}$, representing
the dynamical equations of motion for each component in isolation, and
can equivalently be seen as a generalized impedance or spring
constant; its inverse, $\hat{Y}^{(0)} = \hat{Z}^{(0)-1}$, represents a
generalized admittance for the components in isolation, such that an
external force $\ket{F^{(0)}}$ on the components \emph{in isolation}
produces a total displacement $\ket{x} = \hat{Y}^{(0)}
\ket{F^{(0)}}$. However, energy exchange among components is only
possible if couplings are present: as will become clearer later, these
couplings may act directly between components, or may act through
other components whose equations are eliminated, resulting in
effective self-couplings for the remaining components. These couplings
are generically represented by the linear operator $\Delta\hat{Z}$:
the force $\ket{F}$ on other components due to a displacement
$\ket{x}$ from equilibrium of a given component can be written as
$\ket{F} = -\Delta\hat{Z} \ket{x}$.

We generally assume this system to be reciprocal, so that in this
complex Hilbert space, $\hat{Z}^{(0)} = \hat{Z}^{(0)\top}$ and
$\Delta\hat{Z} = \Delta\hat{Z}^{\top}$ hold, where ${}^{\top}$ denotes
the transpose \emph{without} conjugation and not the Hermitian adjoint
${}^{\dagger}$, and reciprocity of related operators follows from
these relations. Additionally, causality implies that
$\hat{Y}^{(0)}(-\omega^{\star}) = \hat{Y}^{(0)\star}(\omega)$ for any
complex frequency $\omega$; passivity implies that $\asym(Y^{(0)})$,
where $\asym(\hat{A}) \equiv (\hat{A} - \hat{A}^{\dagger})/(2\im)$ for
any operator $\hat{A}$, which represents the dissipative contribution
to the response of the uncoupled components, is Hermitian
positive-definite (in the space of its own support) for any real
positive frequency $\omega$~\cite{KrugerPRB2012}.

We now turn to the equations of motion for this system in the presence
of coupling between its different components. In particular, the total
generalized displacement can be written as the sum of the initial
displacement $\ket{x^{(0)}}$ and the response of the components
$\hat{Y}^{(0)}$, in isolation, to the total generalized force
$\ket{F}$; in order to avoid double-counting various contributions,
the total generalized force simply arises from the total displacements
through the couplings between different components $\Delta\hat{Z}$, as
we assume that no other external forces contribute to the system
dynamics, and these couplings are the only way for energy to be
transmitted among different components. Mathematically, this is
written as
\begin{equation}
  \begin{split}
    \ket{x} &= \ket{x^{(0)}} + \hat{Y}^{(0)} \ket{F} \\
    \ket{F} &= -\Delta\hat{Z} \ket{x}
  \end{split}
\end{equation}
and we formally solve this to yield
\begin{equation} \label{eq:totallinearresponse}
  \ket{x} = \hat{Y} \hat{Z}^{(0)} \ket{x^{(0)}}
\end{equation}
where we define the total response $\hat{Y} = \hat{Z}^{-1}$ in terms
of the total equations of motion (generalized impedance) $\hat{Z} =
\hat{Z}^{(0)} + \Delta\hat{Z}$. We stress that the total response
$\hat{Y}$ satisfies the same reciprocity, causality, and passivity
properties as the decoupled response $\hat{Y}^{(0)}$.

At this point, we specify that the system can be partitioned into $N$
components, with each component labeled $n$ specifying a certain set
of DOFs; the operator $\hat{Z}^{(0)}$ (and also $\hat{Y}^{(0)}$ by
extension) can be written as a block-diagonal matrix as it represents
the equations of motion of each component in the absence of coupling
between components, so if $\hat{P}_{n}$ is a projection into the
subspace supported by the DOFs of component $n$, then each diagonal
block of $\hat{Z}^{(0)}$ is $\hat{Z}^{(0)}_{n} = \hat{P}_{n}
\hat{Z}^{(0)} \hat{P}_{n}$, and likewise each diagonal block of
$\hat{Y}^{(0)}$ is $\hat{Y}^{(0)}_{n} = \hat{P}_{n} \hat{Y}^{(0)}
\hat{P}_{n}$, with $\hat{Y}^{(0)}_{n} = \hat{Z}^{(0)-1}_{n}$. (We note
that $\hat{Z}$ and $\hat{Y}$ will generally not be block-diagonal with
respect to the different components.)

If each component $n$ is maintained independently at a corresponding
temperature $T_{n}$ (uniformly for all of the DOFs constituting that
component), then we may write the frequency-domain
fluctuation--dissipation theorem as
\begin{multline} \label{eq:FDT}
  \hat{P}_{m} \bracket{\ket{x^{(0)} (\omega)} \bra{x^{(0)} (\omega')}}
  \hat{P}_{n} = \\ \frac{2\Pi(\omega, T_{n})}{\omega}
  \asym(\hat{Y}^{(0)}_{n} (\omega)) \delta_{mn} \times
  2\pi\delta(\omega - \omega')
\end{multline}
where $\bracket{\ldots}$ represents the quantum statistical
expectation value; our use of the Planck function $\Pi(\omega, T) =
\frac{\hbar\omega}{2}\coth\left(\frac{\hbar\omega}{2k_{\mathrm{B}}
  T}\right)$ implicitly assumes that all DOFs we consider, when
quantized, obey Bose statistics with no chemical potential, which is
appropriate for EM fields and for coupled mechanical oscillators under
consideration. We also point out that the use of
$\asym(\hat{Y}^{(0)}_{n})$, as opposed to $\asym((\hat{Z}^{(0)}_{n} +
\Delta\hat{Z}_{nn})^{-1})$ if $\hat{P}_{n} \Delta\hat{Z} \hat{P}_{n} =
\Delta\hat{Z}_{nn} \neq 0$, is valid because the former includes
dissipation only within component $n$, whereas the latter may
implicitly include dissipation in other coupled components that have
been eliminated.  In order to compute the heat transfer from component
$m$ to component $n$, we account only for fluctuations in component
$m$, so that $\ket{x^{(0)}} = \hat{P}_{m} \ket{x^{(0)}}$ will hold,
and compute the work done on component $n$ according to $\ket{F} =
-\hat{P}_{n} \Delta\hat{Z} \ket{x}$ where $\ket{x} = \hat{Y}
\hat{Z}^{(0)} \hat{P}_{m} \ket{x^{(0)}}$. We then write the absorbed
power (energy transfer) as $\dot{W} = \int_{-\infty}^{\infty}
\int_{-\infty}^{\infty} \bracket{\trace{-\im\omega \hat{P}_{n}
    \ket{x(\omega)} \bra{F(\omega')} \hat{P}_{n}} e^{-\im(\omega -
    \omega')t}}~\frac{\mathrm{d}\omega~\mathrm{d}\omega'}{(2\pi)^{2}}$.
Algebraic manipulations involving the definitions of $\ket{x}$ and
$\ket{F}$, along with the fluctuation--dissipation theorem
in~\eqref{FDT} applied to $\ket{x^{(0)}}$ and the causality properties
of the relevant response quantities, yield
\begin{multline*}
  \dot{W} = \frac{2}{\pi} \int_{0}^{\infty} \Pi(\omega, T_{m})
  \times \\ \trace{\hat{P}_{m} \asym(\hat{Z}^{(0)\dagger}) \hat{P}_{m}
    \hat{Y}^{\dagger} \asym(\hat{P}_{n} \Delta\hat{Z}) \hat{Y}
    \hat{P}_{m}}~\mathrm{d}\omega
\end{multline*}
as the gross energy transfer from component $m$ at temperature $T_{m}$
to component $n$ among a collection of an arbitrary number of
thermalized components. From this, we define the NEGF energy transfer
spectrum between components $m$ and $n$ as
\begin{equation} \label{eq:Phimn}
  \Phi^{(m)}_{n} = 4~\trace{\hat{P}_{m} \asym(\hat{Z}^{(0)\dagger})
    \hat{P}_{m} \hat{Y}^{\dagger} \asym(\hat{P}_{n} \Delta\hat{Z})
    \hat{Y} \hat{P}_{m}}
\end{equation}
independently of the temperature of each component, while the
integrated net power transfer can be written as
\begin{equation}
  \dot{W}_{m \to n} = \int_{0}^{\infty} (\Pi(\omega, T_{m}) -
  \Pi(\omega, T_{n}))\Phi^{(m)}_{n}~\frac{\mathrm{d}\omega}{2\pi}
\end{equation}
in terms of the component temperatures $T_{m}$ and
$T_{n}$. Furthermore, the heat transfer coefficient (thermal
conductance) between two components may be derived by replacing
$\Pi(\omega, T_{m}) - \Pi(\omega, T_{n})$ in the integrand with
$\lim_{T_{m} \to T_{n}} \frac{\Pi(\omega, T_{m}) - \Pi(\omega,
  T_{n})}{T_{m} - T_{n}} = \frac{\partial}{\partial T} \Pi(\omega,
T)$. It is worth noting that reciprocity, which has not been exploited
in these derivations thus far, is required to show that
$\Phi^{(m)}_{n} = \Phi^{(n)}_{m}$ at each frequency.

The formula for $\Phi^{(m)}_{n}$ in~\eqref{Phimn} is valid for any
number of components maintained at their own uniform temperatures, and
constitutes a generalization of Landauer/Caroli formulas often used to
describe PCHT and RHT~\cite{OteyJQSRT2014, KrugerPRB2012,
  VenkataramPRL2018, JinPRB2019, SegalARPC2016, ZhangPRB2018,
  WangPRE2018, KlocknerPRB2018, SadasivamPRB2017, PaulyNJP2008}. The
most fruitful analogies between RHT and PCHT can be extracted from
consideration of heat transfer between two components, through direct
contact or via contact with a third component. In what follows, we
derive formulas for both situations: the first situation is most
relevant to RHT between two bodies or PCHT combined with RHT between
two bodies in direct contact, while the second situation is most
relevant to PCHT combined with RHT between two bodies via a third
intermediate body (typically a thin interface or a small atomic or
molecular junction), though it can also be applied to formally
deriving expressions for RHT between two bodies.

\subsection{Two components in direct contact}

For two components, labeled 1 or 2, in direct contact with each other,
we may write the operators $\hat{Z}^{(0)}$ and $\Delta \hat{Z}$
describing the equations of motion and couplings among these
components may be written as $2\times 2$ block matrices
\begin{align}
  \hat{Z}^{(0)} &=
  \begin{bmatrix}
    \hat{Z}^{(0)}_{1} & 0 \\
    0 & \hat{Z}^{(0)}_{2}
  \end{bmatrix}
  \\
  \Delta \hat{Z} &=
  \begin{bmatrix}
    \Delta \hat{Z}_{1,1} & \Delta \hat{Z}_{1,2} \\
    \Delta \hat{Z}_{2,1} & \Delta \hat{Z}_{2,2}
  \end{bmatrix}
\end{align}
which in turn implies that
\begin{equation}
  \hat{Y}^{(0)} =
  \begin{bmatrix}
    \hat{Y}^{(0)}_{1} & 0 \\
    0 & \hat{Y}^{(0)}_{2}
  \end{bmatrix}
\end{equation}
must also hold; the existence of nontrivial diagonal and off-diagonal
blocks in $\Delta \hat{Z}$ typically arises from couplings to other
components that are mathematically eliminated in favor of these two
components. Evaluation of the energy transfer spectrum~\eqref{Phimn}
requires inversion of these block matrices of operators, which is
saved for the appendix for the sake of brevity in this section. The
result is written as
\begin{multline}
  \Phi = 4~\operatorname{Tr}\Bigg[\asym(\hat{Z}^{(0)\dagger}_{1})
    (\hat{Z}^{(0)\dagger}_{1} + \Delta \hat{Z}^{\dagger}_{1,1})^{-1}
    \times \\ (\hat{1} - \Delta \hat{Z}^{\dagger}_{2,1}
    (\hat{Z}^{(0)\dagger}_{2} + \Delta \hat{Z}^{\dagger}_{2,2})^{-1}
    \Delta \hat{Z}^{\dagger}_{1,2} (\hat{Z}^{(0)\dagger}_{1} + \Delta
    \hat{Z}^{\dagger}_{1,1})^{-1})^{-1} \times \\ \Delta
    \hat{Z}^{\dagger}_{2,1} (\hat{Z}^{(0)\dagger}_{2} + \Delta
    \hat{Z}^{\dagger}_{2,2})^{-1} \times
    \\ \asym(\hat{Z}^{(0)\dagger}_{2}) (\hat{Z}^{(0)}_{2} + \Delta
    \hat{Z}_{2,2})^{-1} \Delta \hat{Z}_{2,1} \times \\ (\hat{1} -
    (\hat{Z}^{(0)}_{1} + \Delta \hat{Z}_{1,1})^{-1} \Delta
    \hat{Z}_{1,2} (\hat{Z}^{(0)}_{2} + \Delta \hat{Z}_{2,2})^{-1}
    \Delta \hat{Z}_{2,1})^{-1} \times \\ (\hat{Z}^{(0)}_{1} + \Delta
    \hat{Z}_{1,1})^{-1}\Bigg]
\end{multline}
and as $\asym(\hat{Z}^{(0)\dagger}_{1})$ and
$\asym(\hat{Z}^{(0)\dagger}_{2})$ are Hermitian positive-semidefinite
operators with well-defined Hermitian square roots, then the energy
transfer spectrum is nonnegative. Exploiting this further allows for
factorizing $\asym(\hat{Z}^{(0)\dagger}_{n}) =
(\asym(\hat{Z}^{(0)\dagger}_{n})^{1/2})^{2}$ for each component $n \in
\{1, 2\}$, and rearranging the trace allows for writing
\begin{multline} \label{eq:Phi2bodydirect}
  \Phi(\omega) = 4~\Bigg\Vert
  \asym(\hat{Z}^{(0)\dagger}_{2})^{1/2} (\hat{Z}^{(0)}_{2} + \Delta
  \hat{Z}_{2,2})^{-1} \Delta \hat{Z}_{2,1} \times \\ (\hat{1} -
  (\hat{Z}^{(0)}_{1} + \Delta \hat{Z}_{1,1})^{-1} \Delta \hat{Z}_{1,2}
  (\hat{Z}^{(0)}_{2} + \Delta \hat{Z}_{2,2})^{-1} \Delta
  \hat{Z}_{2,1})^{-1} \times \\ (\hat{Z}^{(0)}_{1} + \Delta
  \hat{Z}_{1,1})^{-1} \asym(\hat{Z}^{(0)\dagger}_{1})^{1/2}
  \Bigg\Vert_{\mathrm{F}}^{2}
\end{multline}
where $\left\Vert \hat{A} \right\Vert_{\mathrm{F}} =
\sqrt{\trace{\hat{A}^{\dagger} \hat{A}}}$ is the Frobenius norm. This
is the general NEGF formula for the energy transfer spectrum between
two components in direct contact, in terms of their individual and
mutual responses.

\subsection{Two components coupled via an intermediate component}

We now consider the case of the two components, labeled 1 or 2, which
are coupled only to a third intermediate component, labeled 3, but not
directly to each other. Mathematically, this means the operators
$\hat{Z}^{(0)}$ and $\Delta \hat{Z}$ describing the equations of
motion and couplings among these components may be written as $3\times
3$ block matrices
\begin{align}
  \hat{Z}^{(0)} &=
  \begin{bmatrix}
    \hat{Z}^{(0)}_{1} & 0 & 0 \\
    0 & \hat{Z}^{(0)}_{2} & 0 \\
    0 & 0 & \hat{Z}^{(0)}_{3}
  \end{bmatrix}
  \\
  \Delta \hat{Z} &=
  \begin{bmatrix}
    0 & 0 & \Delta \hat{Z}_{1,3} \\
    0 & 0 & \Delta \hat{Z}_{2,3} \\
    \Delta \hat{Z}_{3,1} & \Delta \hat{Z}_{3,2} & 0
  \end{bmatrix}
\end{align}
which in turn implies that
\begin{equation}
  \hat{Y}^{(0)} =
  \begin{bmatrix}
    \hat{Y}^{(0)}_{1} & 0 & 0 \\
    0 & \hat{Y}^{(0)}_{2} & 0 \\
    0 & 0 & \hat{Y}^{(0)}_{3}
  \end{bmatrix}
\end{equation}
must also hold, where the vanishing of the components
$\Delta\hat{Z}_{1,2} = (\Delta\hat{Z}_{2,1})^{\top}$ follows from the
assumption that components 1 \& 2 have no direct coupling to each
other; we also point out that compared to the general two-component
formula, here we do not include couplings between a given component
and itself (i.e. $\Delta \hat{Z}$ has vanishing diagonal blocks), as
we assume that there are no other components which we have implicitly
eliminated. Once again leaving the details to the appendix, and again
making use of the fact that $\asym(\hat{Y}^{(0)}_{1})$ and
$\asym(\hat{Y}^{(0)}_{2})$ are Hermitian positive-semidefinite
operators to factorize the trace expression, we write~\eqref{Phimn} as
\begin{multline} \label{eq:Phi2bodyviathird}
  \Phi(\omega) = 4~\Bigg\Vert \asym(\hat{Y}^{(0)}_{2})^{1/2} \Delta
  \hat{Z}_{2,3} \times \\ (\hat{Z}^{(0)}_{3} - \Delta \hat{Z}_{3,1}
  \hat{Y}^{(0)}_{1} \Delta \hat{Z}_{1,3} - \Delta \hat{Z}_{3,2}
  \hat{Y}^{(0)}_{2} \Delta \hat{Z}_{2,3})^{-1} \times \\ \Delta
  \hat{Z}_{3,1} \asym(\hat{Y}^{(0)}_{1})^{1/2}
  \Bigg\Vert_{\mathrm{F}}^{2}
\end{multline}
so that all operator products may be evaluated in the space of
component 3. This is the general NEGF formula for the energy transfer
spectrum between two components in contact only with a third in terms
of their individual responses and mutual couplings.

In the previous subsection, it was noted that for heat transfer
between two components that are directly coupled, the couplings
$\Delta \hat{Z}_{mn}$ for $m, n \in \{1, 2\}$ (particularly the
diagonal blocks) often arise from mathematically eliminating another
component to which these two components are coupled, even if those are
the only couplings. At this point, we rigorously prove this
equivalence for the specific case where the two components are
physically coupled only to a third component. We start by
rewriting~\eqref{totallinearresponse} in terms of the degrees of
freedom of the three components and noting that $\ket{x^{(0)}} =
\hat{P}_{1} \ket{x^{(0)}}$ can be used when considering energy
transfer from component 1 to component 2. Explicitly, this means
writing
\begin{equation*}
  \begin{bmatrix}
    \hat{Z}^{(0)}_{1} & 0 & \Delta \hat{Z}_{1,3} \\
    0 & \hat{Z}^{(0)}_{2} & \Delta \hat{Z}_{2,3} \\
    \Delta \hat{Z}_{3,1} & \Delta \hat{Z}_{3,2} & \hat{Z}^{(0)}_{3}
  \end{bmatrix}
  \begin{bmatrix}
    \ket{x_{1}} \\
    \ket{x_{2}} \\
    \ket{x_{3}}
  \end{bmatrix}
  =
  \begin{bmatrix}
    \hat{Z}^{(0)}_{1} \ket{x^{(0)}_{1}} \\
    0 \\
    0
  \end{bmatrix}
\end{equation*}
and then eliminating $\ket{x_{3}} = -\hat{Y}^{(0)}_{3} (\Delta
\hat{Z}_{3,1} \ket{x_{1}} + \Delta \hat{Z}_{3,2} \ket{x_{2}})$. This
yields the simpler equation in terms of $2\times 2$ block matrices
\begin{multline*}
  \begin{bmatrix}
    \hat{Z}^{(0)}_{1} - \Delta \hat{Z}_{1,3} \hat{Y}^{(0)}_{3} \Delta
    \hat{Z}_{3,1} & -\Delta \hat{Z}_{1,3} \hat{Y}^{(0)}_{3} \Delta
    \hat{Z}_{3,2} \\
    -\Delta \hat{Z}_{2,3} \hat{Y}^{(0)}_{3} \Delta \hat{Z}_{3,1} &
    \hat{Z}^{(0)}_{2} - \Delta \hat{Z}_{2,3} \hat{Y}^{(0)}_{3} \Delta
    \hat{Z}_{3,2}
  \end{bmatrix}
  \begin{bmatrix}
    \ket{x_{1}} \\
    \ket{x_{2}}
  \end{bmatrix}
  \\ =
  \begin{bmatrix}
    \hat{Z}^{(0)}_{1} \ket{x^{(0)}_{1}} \\
    0
  \end{bmatrix}
\end{multline*}
whence the replacements
\begin{align*}
  \hat{Z}^{(0)} &\to \begin{bmatrix}
    \hat{Z}^{(0)}_{1} & 0 \\
    0 & \hat{Z}^{(0)}_{2}
  \end{bmatrix} \\
  \Delta \hat{Z} &\to \begin{bmatrix}
    -\Delta \hat{Z}_{1,3} \hat{Y}^{(0)}_{3} \Delta \hat{Z}_{3,1} &
    -\Delta \hat{Z}_{1,3} \hat{Y}^{(0)}_{3} \Delta \hat{Z}_{3,2} \\
    -\Delta \hat{Z}_{2,3} \hat{Y}^{(0)}_{3} \Delta \hat{Z}_{3,1} &
    -\Delta \hat{Z}_{2,3} \hat{Y}^{(0)}_{3} \Delta \hat{Z}_{3,2}
  \end{bmatrix}
\end{align*}
may be made. Hence, the remainder of the derivation of the expression
for heat transfer is the same, as~\eqref{FDT} for component 1 and the
expression $\dot{W} = \int_{-\infty}^{\infty} \int_{-\infty}^{\infty}
\bracket{\trace{-\im\omega \hat{P}_{2} \ket{x(\omega)}
    \bra{F(\omega')} \hat{P}_{2}} e^{-\im(\omega -
    \omega')t}}~\frac{\mathrm{d}\omega~\mathrm{d}\omega'}{(2\pi)^{2}}$
for the power transfer are both unchanged, thereby proving the
equivalence between the two expressions (\Eqref{Phi2bodydirect} and
\Eqref{Phi2bodyviathird}) for the general NEGF energy transfer
spectrum with these identifications in mind.

Writing the energy transfer spectrum as~\eqref{Phi2bodyviathird} can
not only clarify analogies between PCHT and RHT, but it also naturally
leads to expressions for upper bounds on the spectrum. To derive such
bounds, it will be helpful to define the operators $\hat{\Lambda}_{n}
= \Delta \hat{Z}_{n,3}^{\dagger} \asym(\hat{Y}^{(0)}_{n}) \Delta
\hat{Z}_{n,3}$, being the dissipation of each component $n \in \{1,
2\}$ multiplied by the corresponding couplings to component 3, and the
Green's function of component 3
\begin{equation}
  \hat{Y}_{3} \equiv (\hat{Z}^{(0)}_{3} - \Delta \hat{Z}_{3,1}
  \hat{Y}^{(0)}_{1} \Delta \hat{Z}_{1,3} - \Delta \hat{Z}_{3,2}
  \hat{Y}^{(0)}_{2} \Delta \hat{Z}_{2,3})^{-1}
\end{equation}
which is modified from its bare value $\hat{Y}^{(0)}_{3}$ due to
couplings to components 1 \& 2. Given this, we will show that the
energy transfer spectrum can be written in the Landauer
form~\cite{KlocknerPRB2018, KlocknerPRB2017, SongAIPADV2015,
  SegalARPC2016} as $\Phi = \left\lVert \hat{t}
\right\rVert_{\mathrm{F}}^{2}$ where $\hat{t} =
2\hat{\Lambda}_{2}^{1/2} \hat{Y}_{3} \hat{\Lambda}_{1}^{1/2}$. The
goal then will be to place bounds on the eigenvalues of
$\hat{t}^{\dagger} \hat{t}$ at each $\omega$. The fact that
$\hat{t}^{\dagger} \hat{t}$ is Hermitian positive-semidefinite makes
clear that its eigenvalues, called the transmission eigenvalues (as
$\hat{t}^{\dagger} \hat{t}$ is like a transmission intensity matrix),
are all nonnegative, placing a lower bound on their values. The
following will show how to derive upper bounds of unity on the
transmission eigenvalues.

The derivations thus far have actually not made use of the reciprocity
of the system, namely that $\hat{Z}^{(0)}_{3} =
\hat{Z}^{(0)\top}_{3}$, $\hat{Z}^{(0)}_{n} = \hat{Z}^{(0)\top}_{n}$,
and $\Delta \hat{Z}_{n,3} = \Delta \hat{Z}_{3,n}^{\top}$ for $n \in
\{1, 2\}$, but these reciprocity relations are needed for the upper
bounds on the transmission eigenvalues. Additionally, two further
assumptions are needed, namely that $\asym(\hat{Z}^{(0)}_{3}) \to 0$,
and that the block matrices $\Delta \hat{Z}_{n,3}$ for $n \in \{1,
2\}$ are purely real. These assumptions will later be justified for
the particular cases of PCHT as well as RHT.

With this, it can be seen that $\asym(\hat{Y}_{3}) =
\hat{Y}_{3}^{\dagger} \asym(\hat{Y}_{3}^{-1\dagger})
\hat{Y}_{3}$. Expanding the middle term after exploiting
$\asym(\hat{Z}^{(0)}_{3}) = 0$ gives $\asym(\hat{Y}_{3}^{-1}) =
-\hat{\Lambda}_{1} - \hat{\Lambda}_{2}$, as the real-valued and
reciprocal nature of $\Delta \hat{Z}_{n,3}$ imply $\Delta
\hat{Z}_{3,n} \asym(\hat{Y}^{(0)}_{n}) \Delta \hat{Z}_{n,3} = \Delta
\hat{Z}_{n,3}^{\dagger} \asym(\hat{Y}^{(0)}_{n}) \Delta \hat{Z}_{n,3}$
for $n \in \{1, 2\}$. This means $\asym(\hat{Y}_{3}^{-1\dagger}) =
-\asym(\hat{Y}_{3}^{-1}) = \hat{\Lambda}_{1} +
\hat{\Lambda}_{2}$. Therefore, $\asym(\hat{Y}_{3}) =
\hat{Y}_{3}^{\dagger} (\hat{\Lambda}_{1} + \hat{\Lambda}_{2})
\hat{Y}_{3}$. This expression may be rearranged as
$\hat{Y}_{3}^{\dagger} \hat{\Lambda}_{1} \hat{Y}_{3} +
\hat{Y}_{3}^{\dagger} \hat{\Lambda}_{2} \hat{Y}_{3} -
\asym(\hat{Y}_{3}) = 0$, and as $\hat{\Lambda}_{1}$ is Hermitian
positive-semidefinite, then $\hat{\Lambda}_{1}^{1/2}$ exists, so this
expression may be multiplied on the left and right by
$2\hat{\Lambda}_{1}^{1/2}$ to yield $\hat{t}^{\dagger} \hat{t} +
4\hat{\Lambda}_{1}^{1/2} \hat{Y}_{3}^{\dagger} \hat{\Lambda}_{1}
\hat{Y}_{3} \hat{\Lambda}_{1}^{1/2} - \frac{2}{\im}
\left(\hat{\Lambda}_{1}^{1/2} \hat{Y}_{3} \hat{\Lambda}_{1}^{1/2} -
\hat{\Lambda}_{1}^{1/2} \hat{Y}_{3}^{\dagger} \hat{\Lambda}_{1}^{1/2}
\right) = 0$, where as a reminder, $\hat{t} = 2\hat{\Lambda}_{2}^{1/2}
\hat{Y}_{3} \hat{\Lambda}_{1}^{1/2}$. Finally, adding the identity
operator $\hat{1}$ to both sides yields $\hat{t}^{\dagger} \hat{t} +
4\hat{\Lambda}_{1}^{1/2} \hat{Y}_{3}^{\dagger} \hat{\Lambda}_{1}
\hat{Y}_{3} \hat{\Lambda}_{1}^{1/2} - \frac{2}{\im}
\left(\hat{\Lambda}_{1}^{1/2} \hat{Y}_{3} \hat{\Lambda}_{1}^{1/2} -
\hat{\Lambda}_{1}^{1/2} \hat{Y}_{3}^{\dagger} \hat{\Lambda}_{1}^{1/2}
\right) + \hat{1} = \hat{1}$. This expression can be rewritten as
$\hat{t}^{\dagger} \hat{t} + (\hat{1} - \frac{2}{\im}
\hat{\Lambda}_{1}^{1/2} \hat{Y}_{3} \hat{\Lambda}_{1}^{1/2})^{\dagger}
(\hat{1} - \frac{2}{\im} \hat{\Lambda}_{1}^{1/2} \hat{Y}_{3}
\hat{\Lambda}_{1}^{1/2}) = \hat{1}$, showing that $4\hat{t}^{\dagger}
\hat{t}$ is added another Hermitian positive-semidefinite operator to
yield the identity. Therefore, the eigenvalues of $\hat{t}^{\dagger}
\hat{t}$ can never exceed 1, matching the prior
expressions~\cite{Datta1995, PaulyNJP2008, KlocknerPRB2018,
  SadasivamPRB2017}. Additionally, because the operator $\hat{1} -
\frac{2}{\im} \hat{\Lambda}_{1}^{1/2} \hat{Y}_{3}
\hat{\Lambda}_{1}^{1/2}$ is not the zero operator, its rank must be at
least 1, meaning at least one of its eigenvalues must be strictly
positive; in turn, at least one of the eigenvalues of
$\hat{t}^{\dagger} \hat{t}$ must be strictly less than $1$. We stress
that whenever heat transfer between two components that are directly
coupled can be physically equated to heat transfer between the same
two components with effective couplings only via a third (possibly
aggregate) component, these transmission eigenvalue bounds must hold
for that system. Additionally, we expect that even if $\Delta \hat{Z}$
were to have nonzero blocks other than $\Delta \hat{Z}_{n,3}$ (and
their transposes) for components $n \in \{1, 2\}$, which could
represent more general heat transfer between a pair of components
among a collection of $N$ components (for any integer $N \geq 3$) by
virtue of aggregating the other components into an overall third
intermediate component, similar bounds should hold in general, though
we do not prove that statement; put simply, Landauer bounds of unity
should hold for each channel even between two components connected via
a third where each of these components could in principle be connected
to many other components in turn.

\section{Applications to PCHT} \label{sec:PCHT}
The general NEGF formalism and expression for the energy transfer
spectrum~\eqref{Phimn} applies to PCHT among a collection of material
bodies, modeled as effective harmonic oscillators connected to each
other via harmonic short- or long-range couplings, each maintained at
separate uniform temperatures%% ; this harmonic approximation is valid at
%% temperatures $\lesssim 500~\mathrm{K}$, when the spatial dimensions
%% relevant to energy transport between the bodies under consideration
%% are less than the phonon mean free path in the material, and when
%% other tunable anharmonicities are negligible~\cite{SegalARPC2016,
%%   SadasivamPRB2017, KlocknerPRB2017}
. Prior works have typically
focused on PCHT between two large bodies, typically leads acting as
thermal reservoirs, exchanging heat via harmonic coupling through a
third small body in between, typically a molecular junction or a thin
interfacial film; computationally, this has the benefit of allowing
most matrix evaluations to occur in the much smaller space of the
intermediate body as opposed to the larger space of one of the
leads. Given this, in what follows, we derive the equations of motion
for collective atomic oscillations effecting phonons from the
Lagrangian for three bodies, each comprising collections of coupled
oscillators with masses $m_{\alpha a}$, displacements $x_{\alpha ai}$,
and spring couplings $K_{\alpha ai, \beta bj}$ for body labels
$\alpha, \beta \in \{1, 2, 3\}$, atomic labels $a, b, c$ within each
body, and Cartesian indices $i, j, k \in \{x, y, z\}$, with sources
only in body $\alpha = 1$ denoted $x^{(0)}_{1ai}$. Note for comparison
with previous work that bodies 1 and 2, representing infinite
reservoirs (leads), are typically labeled L and R, while body 3,
representing an compact intermediate (central) device, is typically
labeled C. We emphasize that while our derivations focus on the
particular case of two bodies connected to a third in order to make
connections to past work clearer, the correspondence between abstract
linear operators and specific quantities of interest to PCHT is easily
generalized to PCHT among a collection of coupled bodies.

The Lagrangian for this system is written as
\begin{multline}
  2L = \sum_{a,i} m_{1a} (\dot{x}_{1ai} - \dot{x}^{(0)}_{1ai})^{2} -
  \\ \sum_{a,i,a',i'} K_{1ai,1a'i'} (x_{1ai} -
  x^{(0)}_{1ai})(x_{1a'i'} - x^{(0)}_{1a'i'}) - \\ \sum_{a,i,c,k}
  (K_{1ai,3ck} + K_{3ck,1ai})x_{1ai}x_{3ck} + \\ \sum_{b,j} m_{2b}
  \dot{x}_{2bj}^{2} - \sum_{b,j,b',j'} K_{2bj,2b'j'} x_{2bj} x_{2b'j'}
  - \\ \sum_{b,j,c,k} (K_{2bj,3ck} + K_{3ck,2bj})x_{2bj} x_{3ck} +
  \\ \sum_{c,k} m_{3c} \dot{x}_{3ck}^{2} - \sum_{c,k,c',k'}
  K_{3ck,3c'k'} x_{3ck} x_{3c'k'}
\end{multline}
and minimization of the action $S = \int_{-\infty}^{\infty}
L~\mathrm{d}t$ leads to the time domain classical equations of motion
\begin{multline}
  m_{1a} \ddot{x}_{1ai} + \sum_{a',i'} K_{1ai,1a'i'} x_{1a'i'} +
  \sum_{c,k} K_{1ai,3ck} x_{3ck} = \\ m_{1a} \ddot{x}^{(0)}_{1ai} +
  \sum_{a',i'} K_{1ai,1a'i'} x^{(0)}_{1a'i'} \\ m_{2b} \ddot{x}_{2bj}
  + \sum_{b',j'} K_{2bj,2b'j'} x_{2b'j'} + \sum_{c,k} K_{2bj,3ck}
  x_{3ck} = 0 \\ m_{3c} \ddot{x}_{3ck} + \sum_{c',k'} K_{3ck,3c'k'}
  x_{3c'k'} + \\ \sum_{a,i} K_{3ck,1ai} x_{1ai} + \sum_{b,j} K_{3ck,2bj}
  x_{2bj} =0
\end{multline}
for these displacements. In the frequency domain, these become
\begin{multline}
  -\omega^{2} m_{1a} x_{1ai} + \sum_{a',i'} K_{1ai,1a'i'} x_{1a'i'} +
  \sum_{c,k} K_{1ai,3ck} x_{3ck} = \\ -\omega^{2} m_{1a} x^{(0)}_{1ai} +
  \sum_{a',i'} K_{1ai,1a'i'} x^{(0)}_{1a'i'} \\ -\omega^{2} m_{2b}
  x_{2bj} + \sum_{b',j'} K_{2bj,2b'j'} x_{2b'j'} + \sum_{c,k}
  K_{2bj,3ck} x_{3ck} = 0 \\ -\omega^{2} m_{3c} x_{3ck} + \sum_{c',k'}
  K_{3ck,3c'k'} x_{3c'k'} + \\ \sum_{a,i} K_{3ck,1ai} x_{1ai} +
  \sum_{b,j} K_{3ck,2bj} x_{2bj} = 0
\end{multline}
and these equations can be collected into matrix form with vectors
$x_{\alpha}$ and matrices $K_{\alpha\beta}$ and $M_{\alpha}$, upon
which the identifications $\hat{Z}^{(0)}_{\alpha} \to K_{\alpha\alpha}
- \omega^{2} M_{\alpha}$ and $\Delta \hat{Z}_{\alpha\beta} \to (1 -
\delta_{\alpha\beta})K_{\alpha\beta}$ can be made, where
$K_{\alpha\beta} = K_{\beta\alpha}^{\top}$ are real-valued, and
$M_{\alpha}$ are real-valued too; we note that the as the matrices $K$
encode spring constants which multiply \emph{differences} in atomic
positions (i.e. relative displacements) to yield forces, the diagonal
blocks $K_{\alpha\alpha}$ entering $\hat{Z}^{(0)}_{\alpha}$ should
actually include the effects of couplings to other bodies as are
present in the off-diagonal blocks $K_{\alpha\beta}$ for all $\beta
\neq \alpha$, so that all forces are balanced in the equations of
motion. With these replacements, the energy transfer spectrum becomes
\begin{multline}
  \Phi = \\ 4~\operatorname{Tr}\Bigg[K_{3,1} \asym((K_{1,1} -
    \omega^{2} M_{1})^{-1}) K_{1,3} (K_{3,3} - \omega^{2} M_{3} -
    \\ K_{3,1} (K_{1,1} - \omega^{2} M_{1})^{-1} K_{1,3} - K_{3,2}
    (K_{2,2} - \omega^{2} M_{2})^{-1} K_{2,3})^{-1\dagger} \times
    \\ K_{3,2} \asym((K_{2,2} - \omega^{2} M_{2})^{-1}) K_{2,3}
    (K_{3,3} - \omega^{2} M_{3} - \\ K_{3,1} (K_{1,1} - \omega^{2}
    M_{1})^{-1} K_{1,3} - K_{3,2} (K_{2,2} - \omega^{2} M_{2})^{-1}
    K_{2,3})^{-1}\Bigg]
\end{multline}
where the identifications
\begin{equation}
  \hat{Y}^{(0)}_{\alpha} \to g_{\alpha} = (K_{\alpha\alpha} -
  \omega^{2} M_{\alpha})^{-1}
\end{equation}
as the retarded Green's function of lead $\alpha \in
\{1, 2\}$ (with $g_{\alpha}^{\dagger}$ being the advanced Green's
function),
\begin{equation}
  \hat{Y}_{3} \to G = (K_{3,3} - \omega^{2} M_{3} - K_{3,1} g_{1}
  K_{1,3} - K_{3,2} g_{2} K_{2,3})^{-1}
\end{equation}
as the retarded Green's function of the device including connections
to the leads (with $G^{\dagger}$ being the advanced Green's function),
and
\begin{equation}
  \Lambda_{\alpha} = K_{3,\alpha} \asym(g_{\alpha}) K_{\alpha,3}
\end{equation}
for $\alpha \in \{1, 2\}$ being the dissipation terms at the interface
of the device with each lead can immediately be made. Thus, this
general formalism does reproduce the standard Landauer
formula~\cite{MingoPRB2003, KlocknerPRB2018, KlocknerPRB2017,
  SegalARPC2016}
\begin{equation}
  \Phi(\omega) = 4~\trace{\Lambda_{1} G^{\dagger}
    \Lambda_{2} G}
\end{equation}
for phonon heat transport between two leads across a device. Note that
while $K_{\alpha\alpha}$ and $M_{\alpha}$ are real-valued,
$g_{\alpha}$ is complex-valued because inversion of an
infinite-dimensional matrix is made finite-dimensional by considering
propagation of phonons far from the device interface to be equivalent
to energy loss (so $G$ is also complex-valued in turn); alternatively,
if the leads are large but finite, dissipation may be added
heuristically by replacing, including in the definitions of
$g_{\alpha}$, every instance of $-\omega^{2} M_{\alpha}$ with
$-\im\omega B_{\alpha} - \omega^{2} M_{\alpha}$ for $\alpha \in \{1,
2\}$ where the diagonal positive-definite matrices $B_{\alpha}$
represent appropriate dissipation coefficients for the
oscillators. Additionally, the assumptions underlying the derivation
of the upper bound on the transmission eigenvalues hold here, so those
derivations remain valid in this context: all of the $K$ and $M$
matrices are real-valued and reciprocal, and $\asym(K_{3,3} -
\omega^{2} M_{3}) = 0$ because the compact device will not have any
channels for dissipation in the absence of coupling to reservoirs
(leads). Thus, the general NEGF formalism for heat transfer in linear
response systems can be exactly mapped to the specific NEGF formalism
for linear PCHT.

Physically, the harmonic oscillators represent nuclei dressed by
inner-shell electrons, and the couplings represent chemical bonds
between these oscillators, typically computed via density functional
theory and often anisotropic. We again stress that the correspondences
$\hat{Z}^{(0)}_{\alpha} \to K_{\alpha\alpha} - \omega^{2} M_{\alpha}$
and $\Delta \hat{Z}_{\alpha\beta} \to (1 -
\delta_{\alpha\beta})K_{\alpha\beta}$ for PCHT are generally
applicable even beyond the specific case of two bodies coupled only to
a third intermediate body, which allows more general scenarios for
PCHT to be treated using~\eqref{Phimn}; moreover, these derivations do
not assume that the material bodies exhibit any particular geometry or
spatial symmetry properties. %% That said, we note that for harmonic
%% spring couplings, the diagonal blocks $K_{\alpha\alpha}$ for each body
%% $\alpha$ should actually include the effects of couplings to other
%% bodies (if applicable).

\section{Applications to RHT} \label{sec:RHT}
\begin{figure}[h!]
  \centering
  \includegraphics[width=0.85\columnwidth]{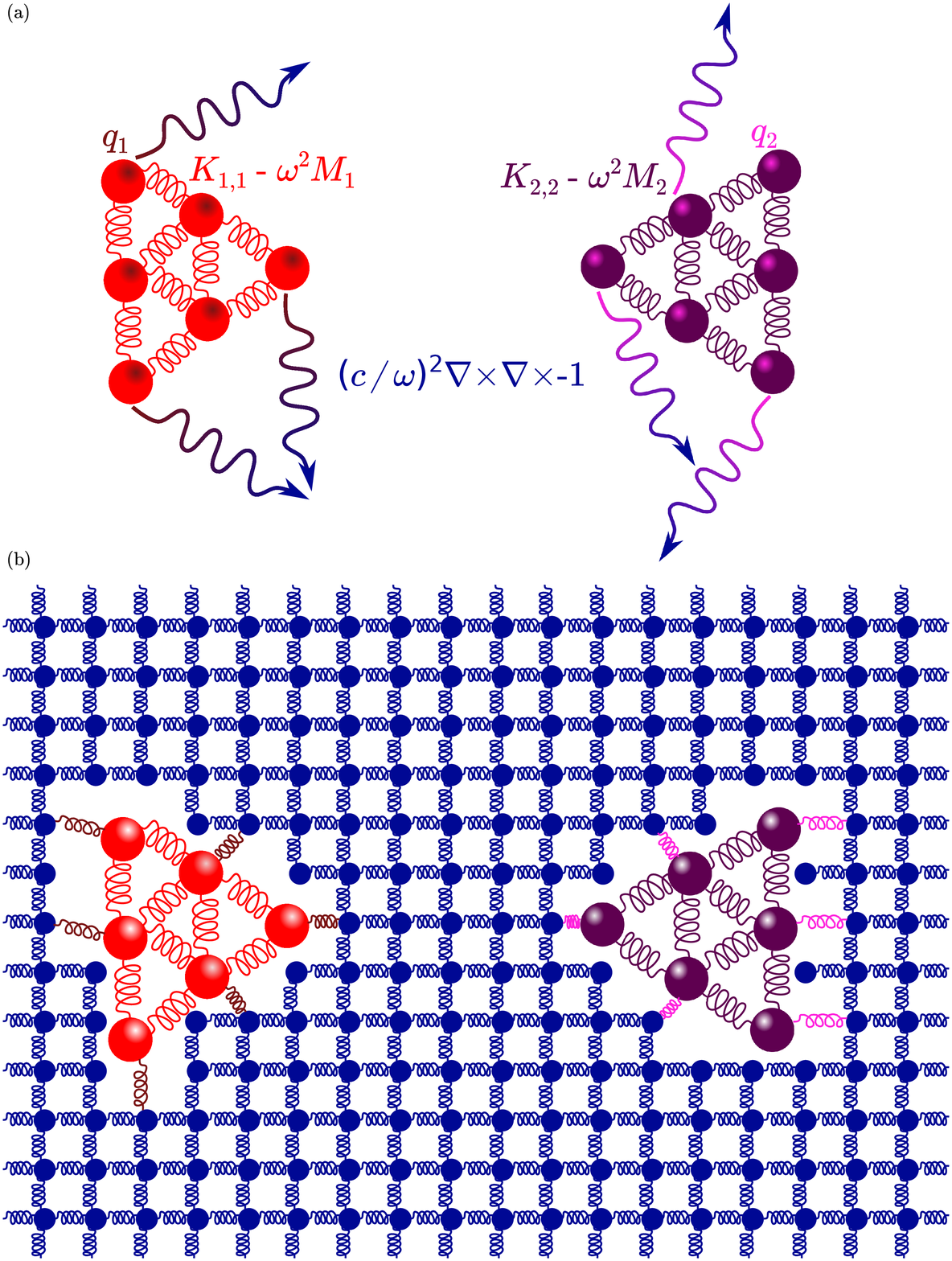}
  \caption{\textbf{Photon radiation.} (a) Radiation of energy in free
    space between compact polarizable bodies. (b) Analogous situation
    for conduction: compact phononic devices are coupled at each atom
    to an infinite lattice supporting propagation of phonons
    infinitely far away.}
  \label{fig:photonradiationschematics}
\end{figure}

The general NEGF formalism and expression for the energy transfer
spectrum~\eqref{Phimn} also applies to RHT among a collection of
linearly polarizable bodies that can radiate EM fields. Prior works
have typically focused on RHT between two polarizable bodies, whether
spatially compact or of infinite extent, in vacuum. The connection to
the above general linear response formalism for heat transfer requires
somewhat more of a conceptual leap compared to the connection for
PCHT. In particular, components 1 \& 2 are the polarizable material
bodies in question, while component 3, rather than representing a
material body, is actually the \emph{vacuum EM field} pervading all of
space. A Lagrangian for this system can easily be written for the case
where the polarizable bodies are made of atomic harmonic oscillators,
with equilibrium positions $\vec{r}_{\alpha a}$ for body $\alpha \in
\{1, 2\}$ and atom label $a$, and with charges $q_{\alpha a}$ that
couple to EM fields; the sources are taken to be in body 1. That said,
the results are generalizable to other linear media whose response
functions are more complicated than those of harmonic oscillators, and
to cases with more than 2 material bodies present; in particular, the
use of partial bound charges associated with harmonic oscillators more
accurately describes polar dielectric media compared to metals, but
the results are generalizable to metals, semimetals, and other media
with susceptibilities that may be nonlocal, inhomogeneous, or
anisotropic.

The Lagrangian for this system is written as
\begin{multline}
  2L = \sum_{a,i} m_{1a} (\dot{x}_{1ai} - \dot{x}^{(0)}_{1ai})^{2} -
  \\ \sum_{a,i,a',i'} K_{1ai,1a'i'} (x_{1ai} -
  x^{(0)}_{1ai})(x_{1a'i'} - x^{(0)}_{1a'i'}) + \\ 2\sum_{a,i} q_{1a}
  \dot{x}_{1ai} A_{i} (\vec{r}_{1a}) + \sum_{b,j} m_{2b}
  \dot{x}_{2bj}^{2} - \\ \sum_{b,j,b',j'} K_{2bj,2b'j'} x_{2bj}
  x_{2b'j'} + 2\sum_{b,j} q_{2b} \dot{x}_{2bj} A_{j} (\vec{r}_{2b})
  + \\ \int \left(\frac{1}{c^{2}} \left(\frac{\partial
    \vec{A}}{\partial t}\right)^{2} - (\nabla \times
  \vec{A})^{2}\right)~\mathrm{d}^{3} x
\end{multline}
introducing the magnetic potential $\vec{A}$, working in the Weyl
gauge (vanishing electric potential). Minimizing the action $S =
\int_{0}^{\infty} L~\mathrm{d}t$ leads to the time domain classical
equations of motion
\begin{multline}
  m_{1a} \ddot{x}_{1ai} + \sum_{a',i'} K_{1ai,1a'i'} x_{1a'i'} -
  q_{1a} E_{i} (\vec{r}_{1a}) = \\ m_{1a} \ddot{x}^{(0)}_{1ai} +
  \sum_{a',i'} K_{1ai,1a'i'} x^{(0)}_{1a'i'} \\ m_{2b} \ddot{x}_{2bj}
  + \sum_{b',j'} K_{2bj,2b'j'} x_{2b'j'} - q_{2b} E_{j} (\vec{r}_{2b})
  = 0 \\ \left(\nabla \times (\nabla \times) + \frac{1}{c^{2}}
  \frac{\partial^{2}}{\partial t^{2}} \right) \vec{E} = \\
  -\frac{1}{c^{2}} \left(\sum_{a} q_{1a} \ddot{\vec{x}}_{1a}
  \delta^{3} (\vec{x} - \vec{r}_{1a}) + \sum_{b} q_{2b}
  \ddot{\vec{x}}_{2b} \delta^{3} (\vec{x} - \vec{r}_{2b}) \right)
\end{multline}
for the displacements $x_{1ai}$ and $x_{2bj}$ and electric field
$\vec{E}(t, \vec{x}) = -\frac{1}{c} \frac{\partial \vec{A}}{\partial
  t}$, where the magnetic contribution to the Lorentz force
$\frac{q_{\alpha a}}{c} \dot{\vec{x}}_{\alpha a} \times \vec{B}$ is
dropped for each atom as it is a nonlinear term that has negligible
contribution for speeds much less than the speed of light $c$ (which
is generally true for thermal fluctuations at reasonable
temperatures). Although the third equation should initially be written
in terms of $\vec{A}$, a partial time derivative is applied to both
sides of the equation to simplify the equations in terms of
$\vec{E}$. In the frequency domain, these equations of motion become
\begin{multline}
  -\omega^{2} m_{1a} x_{1ai} + \sum_{a',i'} K_{1ai,1a'i'} x_{1a'i'}
  - q_{1a} E_{i} (\vec{r}_{1a}) = \\ -\omega^{2} m_{1a} x^{(0)}_{1ai}
  + \sum_{a',i'} K_{1ai,1a'i'} x^{(0)}_{1a'i'} \\
  -\omega^{2} m_{2b} x_{2bj} + \sum_{b',j'} K_{2bj,2b'j'} x_{2b'j'}
  - q_{2b} E_{j} (\vec{r}_{2b}) = 0 \\
  \left(\frac{c^{2}}{\omega^{2}} \nabla \times (\nabla \times) -
  1\right) \vec{E} = \\ \left(\sum_{a} q_{1a} \vec{x}_{1a} \delta^{3}
  (\vec{x} - \vec{r}_{1a}) + \sum_{b} q_{2b} \vec{x}_{2b} \delta^{3}
  (\vec{x} - \vec{r}_{2b}) \right)
\end{multline}
and these equations may again be collected into matrix form and
identified with the generic linear response operators. For polarizable
bodies $\alpha \in \{1, 2\}$, the operators
\begin{equation}
  \hat{Z}^{(0)}_{\alpha} \to K_{\alpha\alpha} - \omega^{2} M_{\alpha}
\end{equation}
are the equations of motion defining the response. Meanwhile,
$\vec{E}$ is a field defined throughout all space, so matrix products
correspond to convolution integrals in space: this means the operators
\begin{align}
  \hat{Z}^{(0)}_{3} &\to \frac{c^{2}}{\omega^{2}} \nabla \times (\nabla \times) - 1 \\
  \hat{Y}^{(0)}_{3} &= \hat{Z}^{(0)-1}_{3} \to \Gvac
\end{align}
correspond to the vacuum Maxwell partial differential operator and
associated Green's function. Finally, in the first, second, and third
equations, the coupling to the third component, i.e. the vacuum EM
field, corresponds to
\begin{equation}
  \Delta \hat{Z}_{\alpha,3} \to -\sum_{a} q_{\alpha a} \delta_{ij}
  \delta^{3} (\vec{x} - \vec{r}_{\alpha a})
\end{equation}
which is the convolution operator representing the charge density of
point dipoles constituting each polarizable body (with a sign flip due
to the convention chosen for the general linear response formulas):
these coupling operators are real-valued reciprocal operators, as
evinced in the equations of motion. This also means that for $\alpha
\in \{1, 2\}$, the material response operators may be written in
position space as
\begin{multline}
  \Delta \hat{Z}_{3,\alpha} \hat{Y}^{(0)}_{\alpha} \Delta
  \hat{Z}_{\alpha,3} \to \VV_{\alpha ij} (\omega, \vec{x},
  \vec{x}') = \\ \sum_{a,a'} q_{\alpha a} ((K_{\alpha\alpha} -
  \omega^{2} M_{\alpha})^{-1})_{ai,a'j} q_{\alpha a'} \delta^{3}
  (\vec{x} - \vec{r}_{\alpha a}) \delta^{3} (\vec{x}' -
  \vec{r}_{\alpha a'})
\end{multline}
which is exactly the susceptibility $\VV_{\alpha}$ of a
collection of point dipolar harmonic oscillators, while
\begin{equation}
  \hat{Y}_{3} \to (\GG^{\mathrm{vac}-1} - \VV_{1} -
  \VV_{2})^{-1} = \GG
\end{equation}
is exactly the Maxwell Green's function in the presence of
susceptibilities $\hat{\chi}_{\alpha}$. Thus, the heat transfer
between the two polarizable bodies can be written as
\begin{equation} \label{eq:GFPhi2bodyRHT}
  \Phi(\omega) = 4~\trace{\asym(\VV_{1})
    \GG^{\dagger} \asym(\VV_{2}) \GG}
\end{equation}
which exactly matches the fluctuational EM
expression~\cite{JinPRB2019}. Additionally, the assumptions underlying
the derivation of the upper bound on the transmission eigenvalues hold
here, so those derivations remain valid in this context: the coupling
operators representing the negative charge densities and real-valued
and reciprocal, and $\asym(\GG^{\mathrm{vac}-1}) = 0$ comes
from the properties of Maxwell's equations, while the fact that
$\asym(\Gvac)$ does not vanish due to free space
supporting outward propagation of EM energy is irrelevant to those
particular derivations. Thus, the general NEGF formalism for heat
transfer in linear response systems can be exactly mapped to the
specific fluctuational EM formalism for linear RHT.

\begin{table*}[t] 
  \centering
  \begin{tabular}{|l|l|l|}
    \hline Heat Transfer Mechanism & Phonons & Photons \\ \hline \hline
    Components 1, 2 & Infinite reservoirs (leads) & Polarizable bodies
    \\ \hline
    Component 3 & Compact central device & Vacuum EM field (all space)
    \\ \hline $\hat{Y}^{(0)}_{\alpha}$: $\alpha \in \{1, 2\}$ &
    \shortstack{Uncoupled lead mechanical Green's function} &
    Susceptibilities $\VV_{\alpha}$ \\ \hline
    $\Delta \hat{Z}_{\alpha,3}$: $\alpha \in \{1, 2\}$ & Interface
    lead/device harmonic couplings & All atom charges \\ \hline
    $\hat{Y}_{3}$ & \shortstack{Coupled device mechanical Green's
      function} & \shortstack{Maxwell Green's function
      $(\GG^{\mathrm{vac}-1} - \VV_{1} -
      \VV_{2})^{-1}$} \\ \hline
  \end{tabular}
  \caption{Comparison of components and relevant linear response
    quantities between PCHT and RHT within our NEGF heat-transfer formalism.}
  \label{tab:comparisontab}
\end{table*}

Physically, the harmonic oscillators may represent valence electrons
or nuclei dressed by inner-shell electrons, and the couplings, namely
the effective charges, along with the effective masses and spring
constants are again computed via density functional theory. We again
stress that the correspondences $\Delta \hat{Z}_{3,\alpha}
\hat{Y}^{(0)}_{\alpha} \Delta \hat{Z}_{\alpha,3} \to
\VV_{\alpha ij} (\omega, \vec{x}, \vec{x}')$ for RHT are
generally applicable even for more than two polarizable bodies coupled
to the vacuum EM field, which allows more general scenarios for PCHT
to be treated using~\eqref{Phimn}~\cite{PolimeridisPRB2015,
  VenkataramPRL2018}. Furthermore, the derivation of Landauer-like
formulas for RHT~\eqref{GFPhi2bodyRHT} is generally applicable for
linear media even when the susceptibilities $\VV_{\alpha}$ do
not describe harmonic oscillator response functions; our use of
harmonic oscillators was for convenience in writing a Lagrangian and
explaining salient features through physical intuition. Finally, we
emphasize that unlike previous work which has typically depended on
high-symmetry geometries and the assumption of the EM near-field
regime~\cite{TangPRAPP2019, WangPRE2018, ZhangPRB2018, JiangPRB2017,
  YuNATURE2017}, these derivations are applicable to arbitrary
geometries from the near- through far-field regimes.

\section{Comparisons between PCHT and RHT} \label{sec:comparison}

Before proceeding, it is useful to summarize the comparisons between
PCHT and RHT specifically focusing on the case of two bodies
interacting through a third component (either a third body for PCHT or
the EM field for RHT), an analogy which is summarized
Table~\ref{tab:comparisontab} and illustrated schematically
in~\figref{photonradiationschematics}. While the basic formalisms are
essentially identical and both obey the same upper bounds, in what
follows we emphasize a few of the distinctions.

The typical situation considered for PCHT involves two semi-infinite
leads connected by a much smaller molecular junction or interfacial
region. As a result, when mapping $\hat{Y}^{(0)}_{\alpha} \to
g_{\alpha}$ for $\alpha \in \{1, 2\}$, even though the microscopic
oscillators have no dissipation so $\asym(g_{\alpha}^{-1}) \to 0$, the
fact that the leads are semi-infinite and act as thermodynamic
reservoirs means $\asym(g_{\alpha}) \neq 0$: this represents loss of
energy through far-field propagation of phonons into the bulks of the
leads. Meanwhile, when mapping $\hat{Y}^{(0)}_{3} \to g_{3}$ for the
junction or interfacial region, the compactness of that intermediate
body precludes dissipation through far-field propagation of phonons,
so not only is it true that $\asym(g_{3}^{-1}) \to 0$ but it is also
true that $\asym(g_{3}) \to 0$. Moreover, the smallness of the
intermediate body means that it is typically easier to evaluate the
matrix products and inverses in the space of the intermediate body
through~\eqref{Phi2bodyviathird}. The situation is flipped for RHT,
where typically energy exchange is considered between two compact
bodies via EM fields that propagate through all of space. As a result,
when mapping the response of lossless oscillators constituting each
polarizable body in the mapping $\Delta \hat{Z}_{3,\alpha}
\hat{Y}^{(0)}_{\alpha} \Delta \hat{Z}_{\alpha,3} \to \VV_{\alpha}$ for
compact bodies $\alpha \in \{1, 2\}$, taking literally the lack of
dissipation would strictly imply that $\asym(\VV_{\alpha}) \to 0$, so
heat transfer \& other fluctuational EM phenomena would not
exist. Realistically, these atomic oscillators are not perfectly
lossless but are subject to losses through scattering and propagation
of energy, which we do not consider here; this can be accounted for by
properly including reservoir DOFs in the Lagrangian and
performing some renormalization like decimation as in the phonon case
for a physically-motivated reservoir, or more typically by
phenomenologically adding an appropriate small imaginary part to some
part of $\VV_{\alpha}$.

Meanwhile, when mapping $\hat{Y}^{(0)}_{3} \to \Gvac$ through all of
space, while it is true that $\asym(\GG^{\mathrm{vac}-1}) \to 0$
allows the same Landauer bounds to hold for RHT as for PCHT, the
ability of free space to support outward propagation of EM energy also
means $\asym(\Gvac) \neq 0$. Moreover, the fact that the polarizable
bodies occupy compact regions in space (as opposed to all of space)
means that it is typically easier to evaluate the matrix products and
inverses in the spaces of the polarizable bodies
through~\eqref{Phi2bodydirect}. In particular, by using the operator
correspondences from the previous section and linking
\eqref{Phi2bodyviathird} to a special case of \eqref{Phi2bodydirect}
as above, it can be shown that \eqref{Phi2bodydirect} exactly
reproduces the T-operator formula for RHT~\cite{KrugerPRB2012}. Along
these lines, we finally note that in PCHT, the off-diagonal block of
the Green's function of component 3 in isolation connecting the
respective atoms coupled to each of the other components, which may be
denoted $P_{3(2)} g_{3} P_{3(1)}$, has a size, and therefore a maximum
rank, that scales as the surface areas of component 3 coupled with
each of the other components. For the case of RHT, the analogous
quantity is $\PP_{2} \Gvac \PP_{1}$, where $\PP_{\alpha}$ is the
projection operator onto the volume of body $\alpha$: this seems to
contrast with the dependence on surface area for PCHT. However, the EM
surface equivalence theorem~\cite{HarringtonJEWA1989,
  RengarajanIEEE2000, ReidPRA2013, RodriguezPRB2013, OteyJQSRT2014,
  SCUFF1} shows that the fields radiated by any volumetric
polarization distribution to the exterior of some fictitious bounding
surface can be exactly reproduced in that exterior region by an
equivalent surface current distribution, which therefore suggests that
the rank of $\PP_{2} \Gvac \PP_{1}$ actually scales with the
\emph{surface} of each body, thereby producing a similar result as for
mechanical waves. The underlying physical reasons are a little
different: the general boundary conditions of EM fields at material
interfaces for radiation contrast with the specific form of coupling
of nearest-neighbor atoms for phonon propagation. That said, the
similarities can be intuitively understood as arising from the similar
physics governing mechanical wave propagation through homogeneous
media as EM wave propagation through vacuum or homogeneous media: the
spring constant matrix $K$ governing mechanical wave propagation
through a medium is essentially a discrete-space analogue of the
$\nabla \times (\nabla \times)$ operator governing EM wave
propagation, and both of these operators are then equated to double
time derivatives of the corresponding field quantities. Finally, we
note that in the concluding remarks, we connect this paper to an
accompanying manuscript that leverages this generic NEGF formalism to
generalize recent bounds on RHT~\cite{MoleskyPRB2020,
  VenkataramPRL2020} to include PCHT: we point out that these bounds
rely heavily on the singular values of the off-diagonal blocks
$\PP_{2} \Gvac \PP_{1}$ in the case of RHT, or $P_{3(2)} g_{3}
P_{3(1)}$ in the case of PCHT.

% TODO: maybe try to discuss more about tighter bounds on PCHT and
% RHT, if possible

\begin{figure*}[ht!]
  \centering
  \includegraphics[width=0.95\textwidth]{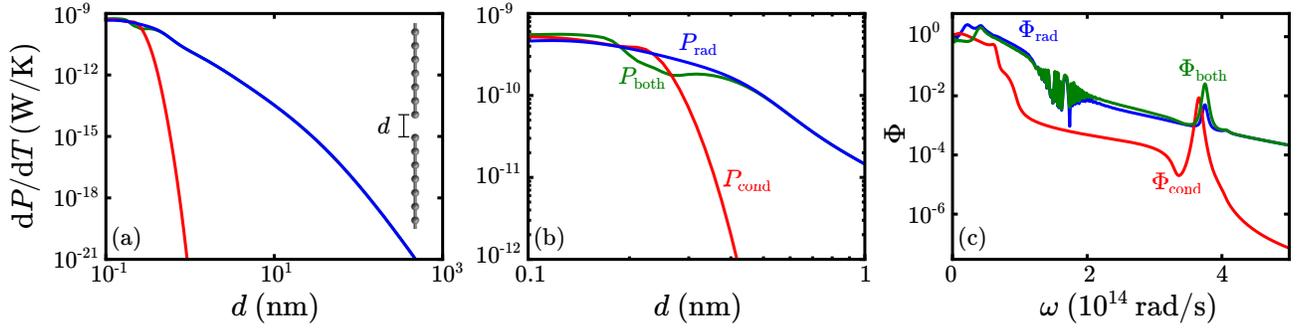}
  \caption{\textbf{Conduction and radiation between collinear wires.}
    (a) Heat transfer coefficient $\frac{\partial P}{\partial T}$ at
    room temperature ($T = 300~\mathrm{K}$) between two collinear 250
    atom-long carbyne wires in vacuum, comparing the cases when heat
    transfer is due purely to radiation (blue) or conduction (red)
    versus both together (green). (b) Same as (a) zoomed in for $d \in
    [0.1~\mathrm{nm}, 1~\mathrm{nm}]$. (c) Landauer energy transfer
    spectrum $\Phi$ (independent of $T$) for $d = 0.281~\mathrm{nm}$,
    clearly demonstrating the existence of nontrivial resonances.}
  \label{fig:carbyne}
\end{figure*}

\section{Unifying PCHT and RHT} \label{sec:unification}
At nanometric and smaller separations, we expect that both PCHT and
RHT could exhibit comparable contributions to overall heat transfer
between two material bodies, whether through approach to direct
contact or through contact with an intermediate
junction~\cite{KimNATURE2015, KloppstechNATURE2017, CuiNATURE2017,
  CuiSCIENCE2017, CuiNATURE2019, StGelaisNATURE2016}. Thus motivated,
we use this section to present a method for unifying both forms of
heat transfer in both of these scenarios. This method is based on the
retarded many-body (RMB) framework of mesoscale fluctuational
EM~\cite{VenkataramPRL2018, VenkataramPRL2017, VenkataramSCIADV2019},
allowing for accurate modeling of fluctuational EM phenomena,
including RHT, in atom-scale systems.

Each body $\alpha \in \{1, 2, 3\}$ comprises $N_{\alpha}$ atoms
labeled $a, b, c$. Each atom is centered at an equilibrium position
$\vec{r}_{\alpha a}$ and has an effective nuclear oscillator of mass
$m_{\mathrm{I}\alpha a}$ which couples to other nuclear oscillators
within the same body and which may couple to nuclear oscillators in
other bodies at interfaces: these couplings are encoded in the
matrices $K_{\mathrm{I}\alpha\alpha}$ within the same body and
$K_{\mathrm{I}\alpha\beta}$ between different bodies, where the former
has dimension $3N_{\alpha} \times 3N_{\alpha}$ while the latter has
dimension $3N_{\alpha} \times 3N_{\beta}$. The effective nuclear
oscillator in each atom is also coupled to an effective valence
electronic oscillator of mass $m_{\mathrm{e}\alpha a}$ through an
isotropic spring constant $k_{\mathrm{e}\alpha a}$. The valence
electronic oscillators couple as point charges to the vacuum EM field
via the charge $q_{\mathrm{e}\alpha a}$; these electrons along with
the inner electrons screen the nuclei, so we model the nuclear
oscillators as having no direct coupling to the EM field. The
displacements of the effective valence electronic oscillators are
labeled $x_{\mathrm{e}\alpha ai}$, while those of the nuclear
oscillators are labeled $x_{\mathrm{I}\alpha ai}$, for Cartesian
direction $i$. We collect the displacements into
$3N_{\alpha}$-dimensional vectors $x_{\mathrm{e}\alpha}$ and
$x_{\mathrm{I}\alpha}$, and the masses, charges, and valence
electronic spring couplings into diagonal $3N_{\alpha} \times
3N_{\alpha}$ matrices $M_{\mathrm{e}\alpha}$, $M_{\mathrm{I}\alpha}$,
$Q_{\mathrm{e}\alpha}$, and $K_{\mathrm{e}\alpha}$. Additionally, the
electric field in vacuum must be evaluated at each equilibrium
position $\vec{r}_{\alpha a}$ when entering the equations of motion
for the effective valence electronic oscillators, so we collect the
$N_{\alpha}$ Cartesian vectors $\vec{E}(\vec{r}_{\alpha a})$ into the
$3N_{\alpha}$-dimensional vector $e_{\mathrm{e}\alpha}$.

For two bodies coming into direct conductive contact (with no third
intermediate material body present) and interacting via the vacuum EM
field, we may use the above matrix notation to
write the equations of motion as
\begin{multline} \label{eq:PCHTRHTequationsofmotion}
  (K_{\mathrm{e}\alpha} - \omega^{2}
  M_{\mathrm{e}\alpha})x_{\mathrm{e}\alpha} - K_{\mathrm{e}\alpha}
  x_{\mathrm{I}\alpha} - Q_{\mathrm{e}\alpha} e_{\mathrm{e}\alpha}
  \\ = ((K_{\mathrm{e}\alpha} - \omega^{2}
  M_{\mathrm{e}\alpha})x_{\mathrm{e}\alpha} - K_{\mathrm{e}\alpha}
  x_{\mathrm{I}\alpha})\delta_{\alpha, 1} \\ (K_{\mathrm{e}\alpha} -
  \omega^{2} M_{\mathrm{I}\alpha}) x_{\mathrm{I}\alpha} + \sum_{\beta}
  K_{\mathrm{I}\alpha\beta} x_{\mathrm{I}\beta} -
  K_{\mathrm{e}\alpha} x_{\mathrm{e}\alpha} = 0
  \\ \left(\frac{c^{2}}{\omega^{2}} \nabla \times (\nabla \times) -
  1\right) \vec{E} = \\ \sum_{\alpha,a} q_{\mathrm{e}\alpha a}
  \vec{x}_{\mathrm{e}\alpha a} \delta^{3} (\vec{x} - \vec{r}_{\alpha
    a})
\end{multline}
for each $\alpha, \beta \in \{1, 2\}$, $a \in \{1, \ldots,
N_{\alpha}\}$, and $i \in \{x, y, z\}$, for sources in body 1. We may
then formally solve the final equation and eliminate
$e_{\mathrm{e}\alpha}$ in favor of $x_{\mathrm{e}\alpha}$ and
$x_{\mathrm{I}\alpha}$, yielding the equations of motion
\begin{multline} \label{eq:PCHTRHTreducedequations}
  (K_{\mathrm{e}\alpha} - \omega^{2}
  M_{\mathrm{e}\alpha})x_{\mathrm{e}\alpha} - K_{\mathrm{e}\alpha}
  x_{\mathrm{I}\alpha} - \sum_{\beta} Q_{\mathrm{e}\alpha}
  G^{\mathrm{vac}}_{\alpha\beta} Q_{\mathrm{e}\beta}
  x_{\mathrm{e}\beta} \\ = ((K_{\mathrm{e}\alpha} - \omega^{2}
  M_{\mathrm{e}\alpha})x^{(0)}_{\mathrm{e}\alpha} -
  K_{\mathrm{e}\alpha} x^{(0)}_{\mathrm{I}\alpha})\delta_{\alpha, 1}
  \\ (K_{\mathrm{e}\alpha} - \omega^{2} M_{\mathrm{I}\alpha})
  x_{\mathrm{I}\alpha} + \sum_{\beta} K_{\mathrm{I}\alpha,\beta}
  x_{\mathrm{I}\beta} - K_{\mathrm{e}\alpha} x_{\mathrm{e}\alpha} = 0
\end{multline}
where $G^{\mathrm{vac}}_{\alpha\beta}$ is the $3N_{\alpha} \times
3N_{\beta}$ matrix whose elements are
$G^{\mathrm{vac}}_{ij}(\omega, \vec{r}_{\alpha a},
\vec{r}_{\beta b})$ for each pair of atomic coordinates. Hence, we
identify the relevant operators as $2\times 2$ block matrices
\begin{equation} \label{eq:PCHTRHToperators}
  \begin{split}
    \hat{Z}^{(0)}_{\alpha} &\to \begin{bmatrix}
      K_{\mathrm{e}\alpha} - \omega^{2} M_{\mathrm{e}\alpha} &
      -K_{\mathrm{e}\alpha} \\
      -K_{\mathrm{e}\alpha} & K_{\mathrm{e}\alpha} +
      K_{\mathrm{I}\alpha\alpha} - \omega^{2} M_{\mathrm{I}\alpha}
    \end{bmatrix} \\
    \Delta \hat{Z}_{\alpha\beta} &\to \begin{bmatrix}
      -Q_{\mathrm{e}\alpha} G^{\mathrm{vac}}_{\alpha\beta}
      Q_{\mathrm{e}\beta} & 0 \\
      0 & K_{\mathrm{I}\alpha\beta} (1 - \delta_{\alpha\beta})
    \end{bmatrix}
  \end{split}
\end{equation}
where the top row and left column blocks represent the effective
valence electronic DOFs, while the bottom row and right
column blocks represent the effective nuclear degrees of
freedom. Strictly speaking, the matrices $-\omega^{2}
M_{\mathrm{e}\alpha}$ and $-\omega^{2} M_{\mathrm{I}\alpha}$ should
respectively be replaced by $-\im\omega B_{\mathrm{e}\alpha} -
\omega^{2} M_{\mathrm{e}\alpha}$ and $-\im\omega B_{\mathrm{I}\alpha}
- \omega^{2} M_{\mathrm{I}\alpha}$ in order to account for nonzero
dissipation, though the dissipation matrices $B_{\mathrm{e}\alpha}$
and $B_{\mathrm{I}\alpha}$ may be taken to be infinitesimal; also,
once again, the diagonal blocks $K_{\mathrm{I}\alpha\alpha}$ entering
$\hat{Z}^{(0)}_{\alpha}$ should actually include the effects of
couplings to nuclear oscillators in other bodies as are present in the
off-diagonal blocks $K_{\mathrm{I}\alpha\beta}$ for all $\beta \neq
\alpha$. With details explained in~\cite{VenkataramPRL2018,
  VenkataramSCIADV2019}, the RMB oscillator matrix parameters
$Q_{\mathrm{e}\alpha}$, $M_{\mathrm{e}\alpha}$,
$M_{\mathrm{I}\alpha}$, $K_{\mathrm{e}\alpha}$, and
$K_{\mathrm{I}\alpha\alpha}$ (the latter initially excluding couplings
to nuclear oscillators in other bodies) along with the equilibrium
atomic positions are all computed using density functional theory
(DFT) for each body in isolation, while the matrices
$B_{\mathrm{e}\alpha}$ and $B_{\mathrm{I}\alpha}$ are assigned
phenomenological values. These $2\times 2$ block matrices can then be
used in place of $\hat{Z}^{(0)}_{\alpha}$ and $\Delta
\hat{Z}_{\alpha\beta}$ in the formula for two components with general
couplings~\eqref{Phi2bodydirect} to find the combined heat transfer
including PCHT and RHT: the couplings among valence electronic and
nuclear DOFs through EM fields means that PCHT and RHT
contributions are not separable, but in fact affect each
other~\cite{WangPRE2018, ZhangPRB2018, KlocknerPRB2017}.

For two bodies whose nuclear coordinates are coupled only to a third
intermediate body, which also has nuclear and valence electronic
DOFs, in which all electronic coordinates are coupled to
the EM field, the formalism is similar to above. In particular, the
formulas in~\eqref{PCHTRHTequationsofmotion} still hold for all bodies
$\alpha, \beta \in \{1, 2, 3\}$, although $K_{\mathrm{I}1,3}$ and
$K_{\mathrm{I}2,3}$ and their transposes are the only nonzero
off-diagonal blocks of $K_{\mathrm{I}}$. With that caveat in mind,
this further means that~\eqref{PCHTRHTreducedequations} and the
correspondences in~\eqref{PCHTRHToperators} holds as well for all
bodies $\alpha, \beta \in \{1, 2, 3\}$. That said, the fact that
$\Delta \hat{Z}_{\alpha\beta}$ has nonzero blocks for all $(\alpha,
\beta)$ means that~\eqref{Phi2bodyviathird} cannot be used. Instead,
the more general formula~\eqref{Phimn} for the energy transfer
spectrum must be used, plugging the $2\times 2$ block matrices
in~\eqref{PCHTRHToperators} into the overall $3\times 3$ block
matrices
\begin{align} 
  \hat{Z}^{(0)} &=
  \begin{bmatrix}
    \hat{Z}^{(0)}_{1} & 0 & 0 \\
    0 & \hat{Z}^{(0)}_{2} & 0 \\
    0 & 0 & \hat{Z}^{(0)}_{3}
  \end{bmatrix}
  \\
  \Delta \hat{Z} &=
  \begin{bmatrix}
    \Delta \hat{Z}_{1,1} & \Delta \hat{Z}_{1,2} & \Delta \hat{Z}_{1,3}
    \\
    \Delta \hat{Z}_{2,1} & \Delta \hat{Z}_{2,2} & \Delta \hat{Z}_{2,3}
    \\
    \Delta \hat{Z}_{3,1} & \Delta \hat{Z}_{3,2} & \Delta \hat{Z}_{3,3}
  \end{bmatrix}
\end{align}
to evaluate~\eqref{Phimn}.

These formulas for the energy transfer spectrum and associated linear
response operators are thus the application of the general NEGF
formalism for combined PCHT and RHT. In contrast to the derivations of
pure RHT which ultimately do not depend on the form of the
susceptibilities $\VV_{\alpha}$ as long as it is linear, these
particular derivations do depend on the harmonicity of the material
models, though they may be generalizable through a more complicated
formalism. However, beyond that approximation as well as the
assumptions regarding material dissipation, these formulas are
independent of specific geometries and material properties, and can be
evaluated in the EM near- or far-field regimes. Additionally, we point
out that unlike previous works which have cast formulas for combined
electronic conduction and RHT in a more complicated (Meir--Wingreen)
form rather than the typical Landauer/Caroli form~\cite{ZhangPRB2018,
  TangPRAPP2019, WangPRE2018} as electrons and photons obey different
quantum statistics, no such complication arises here because phonons
and photons obey the same statistics.

We apply this unified formalism to an illustrative model of heat
transfer between two collinear 250 atom-long atomically thin wires,
taken to be made of carbon (i.e. carbyne wires), and particularly
compute the heat transfer coefficient $\frac{\partial P}{\partial T}$
at room temperature ($T = 300~\mathrm{K}$). Specifically, we compute
the heat transfer coefficient $\frac{\partial
  P_{\mathrm{both}}}{\partial T}$ by calculating the Landauer energy
transfer spectrum $\Phi_{\mathrm{both}}$ arising from
plugging~\eqref{PCHTRHToperators} as written into~\eqref{Phimn},
$\frac{\partial P_{\mathrm{rad}}}{\partial T}$ by computing
$\Phi_{\mathrm{rad}}$ arising from plugging~\eqref{PCHTRHToperators}
with $K_{\mathrm{I}\alpha\beta} = 0$ for $\beta \neq \alpha$ (so
$K_{\mathrm{I}\alpha\alpha}$ refers only to the spring constant
matrices among nuclei for each body in isolation) into~\eqref{Phimn},
and $\frac{\partial P_{\mathrm{cond}}}{\partial T}$ by computing
$\Phi_{\mathrm{cond}}$ arising from plugging~\eqref{PCHTRHToperators}
with $G^{\mathrm{vac}} = 0$ for all pairs of electronic oscillators
into~\eqref{Phimn}; in all cases, $\Phi$ refers to $\Phi^{(1)}_{2}$.
Within each body, as described above, the charges, masses, and spring
constants are all taken from DFT evaluated for each body in isolation,
the matrix elements of the Maxwell Green's function $\Gvac$ are
evaluated in a Gaussian basis to mitigate short-range EM
divergences~\cite{VenkataramPRL2018, HermannCR2017,
  VenkataramPRL2017}, and the dissipation matrices are chosen such
that $B_{\mathrm{e}} = \gamma_{\mathrm{e}} M_{\mathrm{e}}$ \&
$B_{\mathrm{I}} = \gamma_{\mathrm{I}} M_{\mathrm{I}}$ hold with
$\gamma_{\mathrm{e}} = 10^{11}~\mathrm{s}^{-1}$ \&
$\gamma_{\mathrm{I}} = 10^{13}~\mathrm{s}^{-1}$; the damping rates are
chosen phenomenologically to be large enough to allow reasonably
coarse frequency sampling, but small compared to the characteristic
frequencies of the relevant polaritons. For computational simplicity,
these properties are not recomputed as functions of the separation
between the bodies, but while we expect such recomputation to yield
significantly different results due to the greater probability of
supporting longer-wavelength collective electronic and phononic waves
when the wires are in proximity, such recomputation could in principle
be performed consistently with this formalism. Likewise, for
computational simplicity, the off-diagonal blocks of $K_{\mathrm{I}}$
for each body (including both electronic and nuclear oscillator
coordinates) have only the couplings between each end atom nearest to
the other molecule be nonzero, and these are modeled via the Morse
potential, but this could be further generalized in future work. The
Morse potential spring constant for a bond of length $r$ compared to
equilibrium length $a_{0}$ is computed as $k(r) = -\frac{1}{r - a_{0}}
\frac{\partial U_{\mathrm{Morse}}}{\partial r}$, where the potential
energy $U_{\mathrm{Morse}}(r) = U_{\mathrm{min}} (1 -
e^{-\sqrt{k_{0}/(2U_{\mathrm{min}})}(r - a_{0})})^{2}$ exhibits a
harmonic well of depth $U_{\mathrm{min}}$ and curvature defined by the
equilibrium spring constant $k(a_{0}) = k_{0}$, all of which are
empirical parameters, and exponentially decays as $r \gg a_{0}$.

As can be seen in~\figref{carbyne}(a, b), many interesting features
arise from the coupling of conductive and radiative processes. The
exponential decay of the Morse potential with distance means that for
$d > 0.4~\mathrm{nm}$, conduction ceases to have any meaningful effect
on the heat transfer, and the total heat transfer aligns with that of
pure radiation. However, for decreasing $d \leq 0.4~\mathrm{nm}$, not
only does conduction become more significant, but the total heat
transfer including both radiative and conductive processes falls
\emph{below} the corresponding individual cases, and only rises above
both for $d < 0.2~\mathrm{nm}$ before all three powers
saturate. Therefore, this unified formalism is clearly necessary for
subnanometric separations, as the total power including both PCHT and
RHT is not simply the sum of the individual contributions (as has been
found in related systems involving electronic
conduction~\cite{ZhangPRB2018}), but behaves in a much more
complicated way.

In~\figref{carbyne}(c), the Landauer energy transfer spectra $\Phi$
make clear that for small enough $d$ where conduction is nontrivial
(plotted for $d = 0.281~\mathrm{nm}$), the conduction spectrum only
has nontrivial contributions at lower frequencies $\omega <
10^{14}~\mathrm{rad/s}$. Meanwhile, the total spectrum rises above the
radiation spectrum for larger $\omega$ but falls below for smaller
$\omega$: the latter is more relevant given the exponential decay of
$\frac{\partial \Pi(\omega, T)}{\partial T}$ with $\omega$, leading to
$P_{\mathrm{both}} < P_{\mathrm{rad}}$ there. Ultimately, this occurs
due to the confluence of EM screening as captured by the Gaussian
basis functions along with shifts in the response due to conductive
coupling between nuclei of the two different wires: not only does this
shift the frequencies of resonances in the Landauer energy transfer
spectra, but it can also suppress the resulting amplitudes. This
therefore makes clear that the existence of situations where
$P_{\mathrm{both}}$ (or its derivative with respect to $T$) falls
between or below $P_{\mathrm{cond}}$ or $P_{\mathrm{rad}}$ is not
simply a fluke arising from a particular choice of $T$: $\Phi$ is
independent of $T$, yet the spectrum $\Phi_{\mathrm{both}}$, far from
being a simple case of superimposing $\Phi_{\mathrm{rad}}$ on
$\Phi_{\mathrm{cond}}$, shows a delicate interplay among radiative and
conductive effects in creating new hybrid resonances. %% We also point
%% out that these spectra $\Phi$ are consistently less than 10 for all
%% $d$, and are much smaller as $d$ increases further; this contrasts
%% with a Landauer bound of 1500 (equal to the number of degrees of
%% freedom in each molecule, which is the product of the number of atoms
%% and 6, due to 3 Cartesian DOFs for each electronic and
%% nuclear oscillator) for $\Phi$, suggesting the existence of tighter
%% bounds on $\Phi$ at each frequency.
Our calculations are meant to be
qualitatively illustrative of the complexities of heat transfer when
both conduction and radiation contribute: they are not meant to be
quantitatively predictive given the practical limitations in
recomputing relevant oscillator parameters at each separation, but we
stress that these limitations are not fundamental to the formalism we
have presented.

%% That said, practical
%% calculations may remain difficult due to consideration of many
%% different length scales, including the object sizes, the small
%% separations, and the characteristic wavelengths of thermal radiation
%% and conduction. For that reason, while we have presented the formalism
%% here, we do not provide specific examples of such calculations in
%% practice.

\section{Concluding Remarks} \label{sec:conclusion}

We have demonstrated a general NEGF formulation of heat transfer
applicable to a wide variety of bosonic systems. This NEGF framework
is general enough to explain the salient features of PCHT and RHT
separately, show how upper bounds on PCHT can be generalized and then
applied to RHT, and demonstrate how to unify PCHT and RHT in
situations when both are strongly coupled and relevant. The latter is
particularly relevant at atomistic scales or separations, when
continuum material models begin to fail and the net heat transfer is
no longer simply the sum of individual radiative or phononic
contributions. We stress that our approach is general enough to treat
semiclassical heat transfer through other massless bosonic
excitations, not just photons or phonons. Moreover, while our analysis
of combined PCHT and RHT focused on effective valence electronic and
nuclear response as being represented by coupled harmonic oscillators,
more complicated linear response models could be considered as well,
which we leave for future work. We expect this framework to pave the
way for future works investigating the conjunction of PCHT and RHT in
complex geometries, particularly at separations where each is relevant
and where recent experiments have raised questions about where each
form of heat transfer is dominant~\cite{KimNATURE2015,
  KloppstechNATURE2017, CuiNATURE2017, ChiloyanNATURE2015}.

In an accompanying manuscript, we generalize bounds previously derived
for RHT~\cite{MoleskyPRB2020, VenkataramPRL2020} using the generic
NEGF formalism for heat transfer in linear systems presented in this
paper. We particularly apply such bounds to PCHT, showing that
channel-based bounds on PCHT can be much tighter than the Landauer
limits of unity~\cite{BurklePRB2015, KlocknerPRB2017,
  KlocknerPRB2018}.

\emph{Acknowledgments}.---The authors thank Sean Molesky for the
helpful comments and suggestions. This work was supported by the
National Science Foundation under Grants No. DMR-1454836, DMR 1420541,
DGE 1148900, the Cornell Center for Materials Research MRSEC (award
no. DMR1719875), the Defense Advanced Research Projects Agency (DARPA)
under agreement HR00111820046, and the Spanish Ministry of Economy and
Competitiveness (MINECO) (Contract No. FIS2017-84057-P). The views,
opinions and/or findings expressed herein are those of the authors and
should not be interpreted as representing the official views or
policies of any institution.

\appendix
\section{Block matrix inversion for two components coupled directly}
To derive~\eqref{Phi2bodydirect} from~\eqref{Phimn}, the two block
matrices of interest are $\hat{Y} \hat{P}_{1}$ and $\asym(\hat{P}_{2}
\Delta \hat{Z})$, of which the first requires inversion of a block
matrix. In particular, we can immediately evaluate
\begin{equation}
  \asym(\hat{P}_{2} \Delta \hat{Z}) = \begin{bmatrix}
    0 & -\frac{1}{2\im} \Delta \hat{Z}^{\star}_{1,2} \\
    \frac{1}{2\im} \Delta \hat{Z}_{2,1} & \Im(\Delta \hat{Z}_{2,2})
  \end{bmatrix}
\end{equation}
in block form. Meanwhile, standard formulas for inversion of a block
matrix yield
\begin{widetext}
  \begin{equation}
    \hat{Y} \hat{P}_{1} = \begin{bmatrix}
      (\hat{1} - (\hat{Z}^{(0)}_{1} + \Delta \hat{Z}_{1,1})^{-1} \Delta
      \hat{Z}_{1,2} (\hat{Z}^{(0)}_{2} + \Delta \hat{Z}_{2,2})^{-1}
      \Delta \hat{Z}_{2,1})^{-1} (\hat{Z}^{(0)}_{1} + \Delta
      \hat{Z}_{1,1})^{-1} \\
      -(\hat{Z}^{(0)}_{2} + \Delta \hat{Z}_{2,2})^{-1} \Delta
      \hat{Z}_{2,1} (\hat{1} - (\hat{Z}^{(0)}_{1} + \Delta
      \hat{Z}_{1,1})^{-1} \Delta \hat{Z}_{1,2} (\hat{Z}^{(0)}_{2} +
      \Delta \hat{Z}_{2,2})^{-1} \Delta \hat{Z}_{2,1})^{-1}
      (\hat{Z}^{(0)}_{1} + \Delta \hat{Z}_{1,1})^{-1}
    \end{bmatrix}
  \end{equation}
\end{widetext}
where multiplication on the right by $\hat{P}_{1}$ allows for picking
out only the left column block. Putting everything together at this
stage yields

\section{Block matrix inversion for two components coupled only to a third}
To derive~\eqref{Phi2bodyviathird} from~\eqref{Phimn}, the two block
matrices of interest are $\hat{Y} \hat{P}_{1}$ and $\asym(\hat{P}_{2}
\Delta \hat{Z})$, of which the first requires inversion of a block
matrix. To invert the $3\times 3$ block matrices, we exploit the fact
that there are no couplings directly between components 1 \& 2. This
allows for defining subblocks such that
\begin{align}
  \hat{Z}^{(0)} &=
  \begin{bmatrix}
    \hat{Z}^{(0)}_{\mathrm{A}} & 0 \\
    0 & \hat{Z}^{(0)}_{3}
  \end{bmatrix}
  \\
  \Delta \hat{Z} &=
  \begin{bmatrix}
    0 & \Delta \hat{Z}_{\mathrm{A},3} \\
    \Delta \hat{Z}_{3,\mathrm{A}} & 0 \\
  \end{bmatrix}
\end{align}
where aggregate operators for bodies 1 \& 2 are defined as
\begin{align}
  \hat{Z}^{(0)}_{\mathrm{A}} &=
  \begin{bmatrix}
    \hat{Z}^{(0)}_{1} & 0 \\
    0 & \hat{Z}^{(0)}_{2}
  \end{bmatrix}
  \\
  \Delta \hat{Z}_{\mathrm{A},3} &=
  \begin{bmatrix}
    \Delta \hat{Z}_{1,3} \\
    \Delta \hat{Z}_{2,3}
  \end{bmatrix}
  \\
  \Delta \hat{Z}_{3,\mathrm{A}} &= \Delta
  \hat{Z}_{\mathrm{A},3}^{\top}
\end{align}
for this system. Because components 1 \& 2 lie in the top row and left
column blocks of these new $2\times 2$ block matrices, and because
$\hat{Z}^{(0)}$ is block-diagonal in this 2-by-2 aggregate block
representation as well, then $\hat{Z}^{(0)} \hat{P}_{1} = \hat{P}_{1}
\hat{Z}^{(0)} \hat{P}_{1}$, so the energy transfer spectrum $\Phi =
4~\trace{\hat{P}_{1} \asym(\hat{Y}^{(0)}) \hat{P}_{1}
  \hat{Z}^{(0)\dagger} \hat{Y}^{\dagger} \asym(\hat{P}_{2} \Delta
  \hat{Z}) \hat{Y} \hat{Z}^{(0)} \hat{P}_{1}}$ can be computed by
computing $\hat{Y} \hat{P}_{1}$, which is the left block column of
$\hat{Y}$, and $\asym(\hat{P}_{2} \Delta \hat{Z}^{(0)\dagger})$. In
particular, if
\begin{equation}
  \hat{Y} = \begin{bmatrix}
    \hat{Z}^{(0)}_{\mathrm{A}} & \Delta \hat{Z}_{\mathrm{A},3} \\
    \Delta \hat{Z}_{3,\mathrm{A}} & \hat{Z}^{(0)}_{3}
  \end{bmatrix}^{-1}
\end{equation}
then
\begin{widetext}
  \begin{equation}
    \hat{Y} \hat{P}_{1} = \begin{bmatrix}
      \hat{Y}^{(0)}_{\mathrm{A}} \hat{P}_{1} +
      \hat{Y}^{(0)}_{\mathrm{A}} \Delta \hat{Z}_{\mathrm{A},3}
      (\hat{Z}^{(0)}_{3} - \Delta \hat{Z}_{3,\mathrm{A}}
      \hat{Y}^{(0)}_{\mathrm{A}} \Delta \hat{Z}_{\mathrm{A},3})^{-1}
      \Delta \hat{Z}_{3,\mathrm{A}} \hat{Y}^{(0)}_{\mathrm{A}}
      \hat{P}_{1} \\
      -(\hat{Z}^{(0)}_{3} - \Delta \hat{Z}_{3,\mathrm{A}}
      \hat{Y}^{(0)}_{\mathrm{A}} \Delta \hat{Z}_{\mathrm{A},3})^{-1}
      \Delta \hat{Z}_{3,\mathrm{A}} \hat{Y}^{(0)}_{\mathrm{A}}
      \hat{P}_{1}
    \end{bmatrix}
  \end{equation}
\end{widetext}
where $\hat{Y}^{(0)}_{\mathrm{A}} = \hat{Z}^{(0)-1}_{\mathrm{A}}$,
while
\begin{equation}
  \asym(\hat{P}_{2} \Delta \hat{Z}) = \frac{1}{2\im} \begin{bmatrix}
    0 & \hat{P}_{2} \Delta \hat{Z}_{\mathrm{A},3} \\
    -\Delta \hat{Z}_{\mathrm{A},3}^{\dagger} \hat{P}_{2} & 0
  \end{bmatrix}
\end{equation}
is the expression in the block basis.
%% must hold due to reciprocity implying $\Delta \hat{Z}_{3,\mathrm{A}} =
%% \Delta \hat{Z}_{\mathrm{A},3}^{\top}$
Carrying out the operator
product yields the complicated expression
\begin{widetext}
  \begin{multline}
    \hat{P}_{1} \hat{Y}^{\dagger} \asym(\hat{P}_{2} \Delta \hat{Z})
    \hat{Y} \hat{P}_{1} = \frac{1}{2\im} \Bigg(-\hat{P}_{1}
    \hat{Y}^{(0)\dagger}_{\mathrm{A}} \hat{P}_{2} \Delta
    \hat{Z}_{\mathrm{A},3} (\hat{Z}^{(0)}_{3} - \Delta
    \hat{Z}_{3,\mathrm{A}} \hat{Y}^{(0)}_{\mathrm{A}} \Delta
    \hat{Z}_{\mathrm{A},3})^{-1} \Delta \hat{Z}_{3,\mathrm{A}}
    \hat{Y}^{(0)}_{\mathrm{A}} \hat{P}_{1} \\ - \hat{P}_{1}
    \hat{Y}^{(0)\dagger}_{\mathrm{A}} \Delta
    \hat{Z}_{\mathrm{A},3}^{\dagger} (\hat{Z}^{(0)\dagger}_{3} - \Delta
    \hat{Z}_{\mathrm{A},3}^{\dagger} \hat{Y}^{(0)\dagger}_{\mathrm{A}}
    \Delta \hat{Z}_{3,\mathrm{A}}^{\dagger})^{-1} \Delta
    \hat{Z}_{\mathrm{A},3}^{\dagger} \hat{Y}^{(0)\dagger}_{\mathrm{A}}
    \hat{P}_{2} \Delta \hat{Z}_{\mathrm{A},3} (\hat{Z}^{(0)}_{3} -
    \Delta \hat{Z}_{3,\mathrm{A}} \hat{Y}^{(0)}_{\mathrm{A}} \Delta
    \hat{Z}_{\mathrm{A},3})^{-1} \Delta \hat{Z}_{3,\mathrm{A}}
    \hat{Y}^{(0)}_{\mathrm{A}} \hat{P}_{1} \\ + \hat{P}_{1}
    \hat{Y}^{(0)\dagger}_{\mathrm{A}} \Delta
    \hat{Z}_{\mathrm{A},3}^{\dagger} (\hat{Z}^{(0)\dagger}_{3} - \Delta
    \hat{Z}_{\mathrm{A},3}^{\dagger} \hat{Y}^{(0)\dagger}_{\mathrm{A}}
    \Delta \hat{Z}_{3,\mathrm{A}}^{\dagger})^{-1} \Delta
    \hat{Z}_{\mathrm{A},3}^{\dagger} \hat{P}_{2}
    \hat{Y}^{(0)}_{\mathrm{A}} \Delta \hat{Z}_{\mathrm{A},3}
    (\hat{Z}^{(0)}_{3} - \Delta \hat{Z}_{3,\mathrm{A}}
    \hat{Y}^{(0)}_{\mathrm{A}} \Delta \hat{Z}_{\mathrm{A},3})^{-1}
    \Delta \hat{Z}_{3,\mathrm{A}} \hat{Y}^{(0)}_{\mathrm{A}} \hat{P}_{1}
    \\ + \hat{P}_{1} \hat{Y}^{(0)\dagger}_{\mathrm{A}} \Delta
    \hat{Z}_{\mathrm{A},3}^{\dagger} (\hat{Z}^{(0)\dagger}_{3} - \Delta
    \hat{Z}_{\mathrm{A},3}^{\dagger} \hat{Y}^{(0)\dagger}_{\mathrm{A}}
    \Delta \hat{Z}_{3,\mathrm{A}}^{\dagger})^{-1} \Delta
    \hat{Z}_{\mathrm{A},3}^{\dagger} \hat{P}_{2}
    \hat{Y}^{(0)}_{\mathrm{A}} \hat{P}_{1} \Bigg)
  \end{multline}
\end{widetext}
but this can be simplified as follows. The term $\hat{P}_{2}
\hat{Y}^{(0)}_{\mathrm{A}} \hat{P}_{1} = 0$ (and the same is true of
its Hermitian adjoint) because $\hat{Y}^{(0)}_{\mathrm{A}}$ is
block-diagonal, with no correlations between objects 1 \& 2. This
therefore simplifies the expression above to
\begin{multline*}
  \hat{P}_{1} \hat{Y}^{(0)\dagger}_{\mathrm{A}} \Delta
  \hat{Z}_{\mathrm{A},3}^{\dagger} (\hat{Z}^{(0)\dagger}_{3} - \Delta
  \hat{Z}_{\mathrm{A},3}^{\dagger} \hat{Y}^{(0)\dagger}_{\mathrm{A}}
  \Delta \hat{Z}_{3,\mathrm{A}}^{\dagger})^{-1} \times \\ \Delta
  \hat{Z}_{\mathrm{A},3}^{\dagger} \asym(\hat{P}_{2}
  \hat{Y}^{(0)}_{\mathrm{A}}) \Delta \hat{Z}_{\mathrm{A},3} \times \\
  (\hat{Z}^{(0)}_{3} - \Delta \hat{Z}_{3,\mathrm{A}}
  \hat{Y}^{(0)}_{\mathrm{A}} \Delta \hat{Z}_{\mathrm{A},3})^{-1}
  \Delta \hat{Z}_{3,\mathrm{A}} \hat{Y}^{(0)}_{\mathrm{A}} \hat{P}_{1}
\end{multline*}
for which the block-diagonality of $\hat{Y}^{(0)}_{\mathrm{A}}$ once
again allows for writing $\asym(\hat{P}_{2}
\hat{Y}^{(0)}_{\mathrm{A}}) = \hat{P}_{2} \asym(\hat{Y}^{(0)}_{2})
\hat{P}_{2}$. Additionally, performing the block vector-matrix-vector
products within the inverses involving $\hat{Z}^{(0)}_{3}$ gives
$\hat{Z}^{(0)}_{3} - \Delta \hat{Z}_{3,\mathrm{A}}
\hat{Y}^{(0)}_{\mathrm{A}} \Delta \hat{Z}_{\mathrm{A},3} =
\hat{Z}^{(0),3} - \Delta \hat{Z}_{3,1} \hat{Y}^{(0)}_{1} \Delta
\hat{Z}_{1,3} - \Delta \hat{Z}_{3,2} \hat{Y}^{(0)}_{2} \Delta
\hat{Z}_{2,3}$. The terms on the outside, namely
$\hat{Y}^{(0)}_{\mathrm{A}} \hat{P}_{1}$ and its Hermitian adjoint,
can be commuted to yield $\hat{P}_{1} \hat{Y}^{(0)}_{1}$ due to the
block-diagonal structure, and in the trace expression this is then
multiplied on the right by $\hat{Z}^{(0)} \hat{P}_{1}$, the result of
which is simply $\hat{P}_{1}$ as $\hat{Y}^{(0)}_{1} =
\hat{Z}^{(0)-1}_{1}$; this acts to the right of $\Delta
\hat{Z}_{3,\mathrm{A}}$ to yield $\Delta \hat{Z}_{3,1}$. Putting this
all together yields the result in the main text.

\nocite{apsrev41Control} \bibliographystyle{apsrev4-1}

\bibliography{mechanicalpaper}
\end{document}